\documentclass[aps,physrev,superscriptaddress]{revtex4-2}

\usepackage{amsmath}

\usepackage{graphicx}
\usepackage{dcolumn}
\usepackage{bm}

\usepackage[utf8]{inputenc}
\usepackage[T1]{fontenc}
\usepackage{mathptmx}
\usepackage{etoolbox}

\usepackage{siunitx}
\usepackage[percent]{overpic}

\newcommand{\FE}{\mathrm{FE}} 
\newcommand{\FFE}{\mathrm{FFE}} 
\newcommand{\In}{\mathrm{in}} 
\newcommand{\atm}{\mathrm{atm}} 
\newcommand{\hoz}{\mathrm{hoz}} 
\newcommand{\ver }{\mathrm{vert}} 

\newcommand{\OmegaMouth}{\Omega_{m}}
\newcommand{\lambdavec}{\boldsymbol{\lambda}} 
\newcommand{\nvec}{\boldsymbol{n}} 
\newcommand{\qvec}{\boldsymbol{q}} 
\newcommand{\uvec}{\boldsymbol{u}} 
\newcommand{\Uvec}{\boldsymbol{U}} 
\newcommand{\bnabla}{\boldsymbol{\nabla}}
\newcommand{\bcdot}{\boldsymbol{\cdot}}
\newcommand{\btimes}{\boldsymbol{\times}}
\newcommand{\Rey}{\mathrm{Re}}
\newcommand{\LgthScle}{\hat{b}}
\newcommand{\VelScle}{\hat{w}^{*}}

\begin{document}


\title[Mathematical modelling face coverings for virus protection]{Mathematical modelling of face coverings for virus protection}


\author{M. D. Shirley}
\email{matthew.shirley@maths.ox.ac.uk}
\affiliation{ 
Mathematical Institute, Andrew Wiles Building, Woodstock Road, Oxford, OX2 6GG, UK}
 \author{I. M. Griffiths}
\affiliation{ 
Mathematical Institute, Andrew Wiles Building, Woodstock Road, Oxford, OX2 6GG, UK}
\author{J. Houghton}
\affiliation{Virustatic Development Limited, Albion Wharf, 19 Albion Street, Manchester, M1 5LN, UK}
\author{L. Hope}
\affiliation{Virustatic Development Limited, Albion Wharf, 19 Albion Street, Manchester, M1 5LN, UK}
\author{P. Hope}
\affiliation{Virustatic Development Limited, Albion Wharf, 19 Albion Street, Manchester, M1 5LN, UK}
\author{C. J. W. Breward}
\affiliation{ 
Mathematical Institute, Andrew Wiles Building, Woodstock Road, Oxford, OX2 6GG, UK}



\date{\today}

\begin{abstract}

Traditional face masks used extensively, for example, during the COVID-19 pandemic, are relatively impermeable, forcing aerosol-containing breath to flow into the surrounding atmosphere by ``leaking" out from the lateral edges of the mask. This protects close-proximity persons but results in a high viral load being delivered into the air. An alternative face covering, the {\it Virustatic Shield}$^\textrm{\textregistered}$, developed by Virustatic, involves a much more permeable material coated in an active agent, with the goal of increasing the transmission through the mask and thus reducing the viral load released.
In this paper, we build and solve a model that describes the air flow and aerosol transport through a porous face mask. Our model for air flow is based on a low-Reynolds-number approximation to the Navier--Stokes equations, and reduces to a two-dimensional scalar partial differential equation for the local flux through the mask. We use the model to explore the interplay between leakage from the mask sides and flow directly through. We compare and contrast the leakage for traditional face masks and for neck gaiters, and find that neck gaiters have a lower leakage. We utilise a simple model for aerosol transport that balances advection with capture by the mask, and find that there is an ideal permeability, which optimises real-world protection offered by the mask.
\end{abstract}


\maketitle

\section{Introduction}

During the COVID-19 pandemic, the use of face coverings greatly expanded \cite{greenhalgh2024} and evidence shows that they reduced the spread of COVID-19 and other respiratory diseases \cite{feng2020rational,Li2021face,Tracht2010}. Face coverings may be broadly divided into three categories \cite{Schmitt2022review}. The highest protection is offered by respiratory protection devices (respirators), designed to provide a perfect seal between the covering and the face. Such devices are intended to protect the wearer from infection in high-risk environments such as COVID wards in hospitals, and require fit tests by each worker to ensure a particular design fits their face. Respirators are governed by regulations specifying a minimum filtration efficiency and a maximum allowed total inward leakage of air \cite{Schmitt2022review,greenhalgh2024}. Surgical (or medical) face masks offer the next highest level of protection and are also subject to regulated standards. These masks  were originally intended for source control of large droplets, and so no specification of the amount of either inward or outward leakage is given \cite{Schmitt2022review,greenhalgh2024}. Finally, community face coverings (cloth masks, neck gaiters) grew in popularity in the wider population during the pandemic \cite{greenhalgh2024,Schmitt2022review,Lindsley2021}. Regulations for community face coverings were introduced in 2020 due to the pandemic and are again primarily focused on source control, and to a lesser extent on the wearer’s protection \cite{Schmitt2022review}. However, evidence  exists demonstrating that these masks leak a significant proportion of the exhaled air and particles \cite{Viola2021face,okelly2021}. 

There has been considerable interest in understanding and characterising the key factors in the design and use of face coverings to optimise the protection they offer to the wearer and others.
To quantify the performance of face masks, studies have focused on four key metrics, two of which relate to the material and two to the real-world performance \cite{Mittal2023Review}. The properties of the mask material are measured by the \textit{permeability} (often reported as the pressure drop), which quantifies the ability of air to flow through the mask material, and the \emph{filtration efficiency} given by the proportion of particles entering the mask material that are captured. The real-world performance of face masks is quantified by the \textit{leakage ratio}, $\eta$, the proportion of air exhaled that does not pass through the mask material, and the \textit{face-fitted filtration efficiency}, $\FFE$, which is the proportion of exhaled particles captured by the face covering. Note that these definitions can be naturally altered to describe the amount of inward protection offered to the wearer, but we will focus on exhalation in this paper. 

The growth in popularity of community face coverings led to a great deal of innovation, for example, by Virustatic developing the Virustatic Shield$^\textrm{\textregistered}$. Virustatic's aim was to design a face covering, based on a neck gaiter, which was more permeable than a traditional cloth mask, and made of a face conforming elastic material, with the aim of reducing the leakage. This ensures a greater proportion of exhaled air is filtered through the face covering, which is treated with an active coating to maintain a high filtration efficiency \cite{VirustaticMaterial2021}. However, as the regulations intended for surgical masks focus only on material properties, and do not account for leakage, there is no way for their design to progress up the regulatory pathway, despite aiming to offer better real-world protection than a traditional face mask.

There have been a  number of experimental studies on the effect of face coverings worn by human volunteers, including \cite{Viola2021face,Bourrianne2021} who measured the effect of face masks on the distance jets of air travel from the wearer during talking, breathing, and coughing. The measurements presented in \cite{Bourrianne2021} indicate that the velocity of air just as it leaks out the top of the mask is \SI{0.1}{m.s^{-1}} during normal respiration.  In \cite{Duncan2021}, direct measurements of the filtration efficiency and inward face-fitted filtration efficiency of coverings are presented for reusable fabric masks and disposable surgical and N95 masks. They find a variation in the filtration efficiency between the different materials, and that ensuring an effective seal to minimise leakage has a much greater effect on protecting the wearer than the material choice. Other papers \cite{Si2024visualization,pan2021inward,Lindsley2021,kolewe2020,Chiera2022} have focused on using headforms to improve reproducibility. Schlieren imaging is used in \cite{Si2024visualization} to visualise the jets from headforms wearing face coverings to identify where leakage occurs and see the main leakage jets out of the top and sides of the mask. The filtration efficiency and inward and outward face-fitted filtration efficiencies of
eleven different face coverings worn on a mannequin were measured in \cite{pan2021inward}. They found the difference between the material and face-fitted value was greatest for loose-fitting coverings. In \cite{Lindsley2021}, the outwards face-fitted efficiency during a simulated coughing cycle was measured for a variety of face masks, a neck gaiter, and a face shield worn by a mannequin. They found no significant statistical difference between a cloth mask, a surgical mask, and a neck gaiter, but all performed significantly better than the face shield. In \cite{kolewe2020},  a headform was also used to measure the aerosol concentration for a surgical mask, an N-95 mask, and a homemade cloth mask, and found that all face coverings reduced the concentration of aerosols in front of the wearer, but increased it above and to the sides, indicating leakage of particles was occurring. They also found that the differences in concentration between different masks decreased with distance from the wearer. In \cite{Pushpawela2023}, the leakage flow was estimated to be 25\% for cloth and surgical masks, whereas \cite{Chiera2022} estimate a leakage between 43\% and 95\% at low flow rates, with a decrease of about 10\% if the flow rate increases.  These were calculated in both papers by (i) measuring the pressure drop across the face covering in the case where there is leakage, and (ii) when there is a perfect seal, then used an empirical model to estimate the leakage. These results show the wide range of values for leakage available and illustrate how it appears to be sensitive to a particular wearer's facial geometry.

Several studies have measured the permeability and filtration efficiency of a wide range of materials used in the manufacture of cloth and surgical masks   \cite{Zangmeister2020filtration,Duncan2021,konda2020,Lee2020reusable}. Analysis of these results has demonstrated that no simple relationship exists between permeability and filtration efficiency due to the many factors affecting them, including porosity, weave pattern, thread size, and electrostatic properties \cite{Solano2022}. These measurements have been used to parametrise a variety of studies modelling the airflow and aerosol transport of the respiration of a face covering wearer, which then allow the more relevant leakage ratio and face-fitted filtration efficiency to be computed. The models can broadly be divided into two categories: \textit{empirical models} and \textit{computational fluid dynamics} (CFD) models, which we discuss separately.

The simplest empirical model is an equivalent circuit model with a resistor representing the face mask in parallel with a resistor representing flow through a gap \cite{Peric2020,Xu2021}. The resistance through the gap is governed by Darcy's law, while the resistance to flow through the gap is given by Bernoulli's equation \cite{Xu2021}, or a resistor governed by Bernoulli's equation in series with another given by the Hagen--Poiseuille equation 
\cite{Peric2020}. Both models are used to predict the amount of leakage out of the sides of the mask and show that this may be a significant fraction of the total flux. The solution presented in \cite{Peric2020} is validated in two dimensions against the numerical solution of the Navier--Stokes equations and found good agreement in the proportion of leakage predicted.  The model in \cite{Peric2020} has been extended in \cite{Ni2023} to contain additional spatial information by dividing the boundary into segments, represented by parallel resistors, with the resistance in each segment varying according to simulations of the gap width on the perimeter of a face mask. Leakage was again found to be significant. The basic two-resistor model has also been coupled to an equivalent circuit model for the human respiratory system \cite{de2024asymmetric}. Here, it was found that the material of a surgical or N95 mask provided a sufficiently high resistance to breathing that it would become difficult to do so without the effect of leakage lowering the pressure required. Two-resistor equivalent circuit models have been used to quantify the amount of leakage during experiments on mannequin heads in \cite{Pushpawela2023} and \cite{Chiera2022}.   The common feature of all these models is that they admit analytic or fast numerical solutions, allowing for rapid parameter sweeps. However, the equivalent circuit models assume a uniform pressure within the gap behind the mask, making it impossible to predict spatial variation in the flux through the mask.

The other existing modelling approach involves computational fluid dynamics (CFD) simulations, where the full Navier--Stokes equations are solved in realistic three-dimensional geometries representing the face covering, the wearer's head, and the surrounding air. These models have been used to simulate the behaviour during normal breathing \cite{Hariharan2021,Xu2021,xi2022inspiratory,jia2023}, exercise \cite{Hariharan2021}, and coughing \cite{Solano2022,Dbouk2020respiratory,jia2023}. While \cite{Hariharan2021,Xu2021} consider laminar flow models, the remainder all include a turbulence model to account for the high Reynolds number of the flow. The different models are reviewed in \cite{jia2023}, along with a discussion of numerical issues that arise when implementing these models. The porous face covering is either modelled as a separate region
\cite{Hariharan2021,Xu2021} or as a spatially varying body force in the fluid momentum equations~\cite{Dbouk2020respiratory,jia2023,Khosronejad2020,xi2022inspiratory,Solano2022}. Resistance to flow through the mask is commonly assumed to be linear, that is, to obey Darcy's law, for example, in \cite{Hariharan2021,Xu2021,xi2022inspiratory,Solano2021}. However, in \cite{Dbouk2020respiratory,jia2023,Khosronejad2020}, they assume a nonlinear resistance, motivated by experimental evidence that the pressure drop--flux relationship may become quadratic at high flow rates~\cite{Bourrianne2021}.  Given a solution for the airflow, different methods are adopted to estimate the resulting face-fitted filtration efficiency. The simplest model, adopted by \cite{Solano2022,Xu2021}, assumes perfect adherence of the particles to the streamlines, i.e., an advective model, an assumption which is only true for aerosol-sized droplets (diameter less than \SI{5}{\micro m}) \cite{Solano2022}. To simulate larger particles, their inertia must be accounted for, which \cite{Hariharan2021,Dbouk2020respiratory} do using  Lagrangian particle dynamics. In particular, \cite{Dbouk2020respiratory} show that a large proportion of larger particles will impact the surface of a face mask even when the leakage ratio is high. A convection--diffusion model is considered in \cite{Khosronejad2020}. These simulations all provide a large amount of detail about features in the flow, such as the strength of secondary jets and the distance particles may be transmitted from the wearer. However, the computational resources required to compute the fluid flow limit the results reported to a small number of simulations, with \cite{Khosronejad2020} reporting that five high-fidelity simulations of a single coughing cycle required, on average, four months of CPU time on a 1000 CPU cluster, while \cite{Dbouk2020respiratory} reported using three days on a 32 CPU cluster to simulate 10 coughing cycles. The parameter sweep of different mask permeabilities in \cite{Solano2022} required two million CPU hours of computational resources from the National Science Foundation's XSEDE \cite{XSEDEdataset}.

In this paper, we build and solve a simple mathematical model for the exhalation airflow in the gap between a face covering and the wearer's face, and through the face covering. We will use asymptotic analysis to systematically reduce the Navier--Stokes equations and Darcy's law to obtain a reduced model. Our reduced model has the advantage of satisfying physical laws, giving interpretability, and allowing fine spatial resolution, but may be solved in a similar computational time as an empirical model. Using our reduced model, we explore how the amount of air leaking out of the mask depends on the permeability of the face covering, the geometry of the mouth, and the gap between the face covering and the wearer's face. We will also compare two designs of face covering: the standard \textit{face mask}, representing surgical and cotton masks, and a \textit{neck gaiter}. The former is open on all sides and attached to the wearer's face by elastic straps around the ears. The latter comprises a closed loop of fabric that wraps around the entire head, and so is only open to the air at the top and bottom. The neck gaiter is held to the wearer's face by the elasticity of the material itself.  We couple the flow model to a simple advection-capture model for aerosolised viral particles, which we will use to predict face-fitted filtration efficiency. By proposing a simple model linking the particle capture rate to the permeability of the mask material, we will explore the trade-offs in selecting a material to minimise leakage or maximise filtration efficiency.

The remainder of the paper is structured as follows. In Section \ref{sec:Model} we present and nondimensionalise a mathematical model for the airflow and aerosol transport through a face covering and in the gap between the face covering and the wearer's face. In Section \ref{sec:Asymptotics} we will exploit the small size of (i) the aspect ratio of the gap between the face covering and the mouth, and (ii) the reduced Reynolds number, to obtain a reduced problem consisting of a single scalar partial differential equation (PDE) for the flux through the mask at each point. In Section \ref{sec:Leakage} we solve this reduced model explicitly for a constant gap width, and we simulate the model in the case where the gap has variable size. Using these solutions we will explore how the leakage ratio varies with the properties of the material, the size of the gap between the face and the mask, and between a face mask and a neck gaiter. In Section \ref{sec:Efficiency} we will solve the aerosol transport problem and calculate the face-fitted filtration efficiency of a face covering and explore how the parameters in the problem can be varied to optimise this. We present our conclusions in Section \ref{sec:Discussion}.

\section{Mathematical model}\label{sec:Model}

\subsection{Airflow modelling}
\begin{figure}
    \centering
    \includegraphics[width=0.8\linewidth]{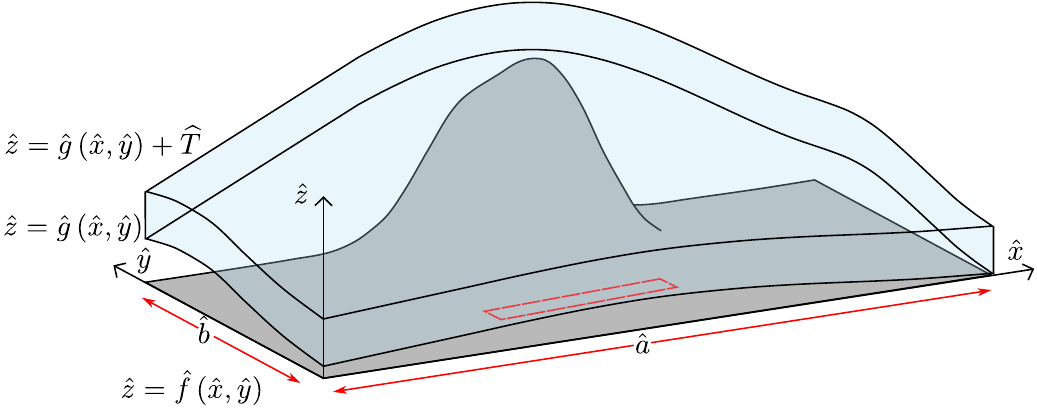}
    \caption{The simplified Cartesian representation considered in this paper. The surface of the face $\hat{z}=\hat{f}\left(\hat{x},\hat{y}\right)$ is shown in dark grey, the face mask occupying the region $\hat{g}\left(\hat{x},\hat{y}\right)<\hat{z}<\hat{g}\left(\hat{x},\hat{y}\right)+\hat{T}$ is shown in light blue. The boundary of the mouth is shown in red.}
    \label{fig:3Ddiagram}
\end{figure}

We begin by constructing a model for the airflow through the gap between the face covering and the wearer's face, and through the face covering. For simplicity, we pose the model only for exhalation, but discuss its modification to inhalation in Section~\ref{sec:Discussion}. We first consider the geometry shown in  Fig.~\ref{fig:3Ddiagram}, where we assume the curvature of the face-covering-wearer's head is negligible, so that the features of their face can be described by a function of the form $\hat{z}=\hat{f}\left(\hat{x},\hat{y}\right)$, where $\left(\hat{x},\hat{y},\hat{z}\right)$ form a three-dimensional Cartesian coordinate system.  We assume that the rear of the face covering is positioned at $\hat{z}=\hat{g}\left(\hat{x},\hat{y}\right)$, so that the width of the gap between the face and the face covering is given by $\hat{h}=\hat{g}-\hat{f}$. We assume that the face covering is inextensible and has constant thickness $\hat{T}$, so that the front of the face covering is located at $\hat{z}=\hat{g}\left(\hat{x},\hat{y}\right)+\hat{T}/({\bf n}_g\cdot{\bf k})$, where ${\bf n}_g$ is the unit normal to the surface at $z=g$ pointing into the mask and $\bf k$ is a vector pointing in the $z$-direction. We further assume that the mask is rectangular in plane view, so that it occupies the region $[0,\hat{a}]\times[0,\hat{b}]$ in the $\hat{x}\hat{y}$-plane, as shown in Fig.~\ref{fig:3Ddiagram}. 

We assume that the air is a Newtonian fluid.  Thus, in the gap between the face and the face covering, fluid velocity, $\hat{\uvec}=\left(\hat{u},\hat{v},\hat{w} \right)$, and pressure, $\hat{p}$, satisfy  the Navier--Stokes equations
\begin{align}
    \hat{\bnabla} \bcdot \hat{\uvec} &=     0,\label{eq:Dim_NS_mass}\\
    \hat{\rho}\left( \frac{\partial \hat{\uvec}}{\partial \hat{t}} +\hat{\uvec}  \bcdot \hat{\bnabla}\hat{\uvec} \right) &= - \hat{\bnabla} \hat{p}  + \hat{\mu} \hat{\bnabla}^2 \hat{\uvec},\label{eq:Dim_NS_momentum}
\end{align}
where $\hat{\rho}$ is the density, and $\hat{\mu}$ the dynamic viscosity of the air, both assumed constant.

We model the flow through the face covering using Darcy's law, which has been shown experimentally to be valid for face coverings at low breathing rates \cite{Bourrianne2021}.  For simplicity, we will assume that the material of the face covering is homogeneous and isotropic. Thus, the governing equations that describe the  (local average) velocity of fluid within the pore-space, $\hat{\Uvec}=\left( \hat{U},\hat{V},\hat{W}\right)$, and the pressure, $\hat{P}$, are
\begin{align}
    \hat{\bnabla}\bcdot\left( \phi \hat{\Uvec}\right) & = 0,\label{eq:Dim_Mask_mass}\\
    \phi \hat{\Uvec} &= - \frac{\hat{K}}{\hat{\mu}} \hat{\bnabla}\hat{P},\label{eq:Dim_Mask_Darcy}
\end{align}
where
$\phi$ is the porosity of the mask, and $\hat{K}$ is the permeability of the mask, both assumed constant.  
We will assume that there is a constant unidirectional flux out of the mouth, in the normal direction to the surface defining the face, and no-flux and no-slip conditions hold on the remainder of the face.  These conditions may be written compactly as 
\begin{equation}
    \hat{\uvec} \times \nvec_{f} = \boldsymbol{0}, \quad  \hat{\uvec} \bcdot \nvec_{f} = \hat{w}_{\In}\left(\hat{x},\hat{y}\right) ~ \text{ at } \hat{z} = \hat{f}\left(\hat{x},\hat{y}\right),\label{eq:DimBCS_face}
\end{equation}
where $\nvec_{f}$ is the unit normal to $\hat{f}$ pointing out of the wearer's face, and
\begin{equation}
    \hat{w}_{\In}\left(\hat{x},\hat{y}\right) = 
    \begin{cases}
         \VelScle& \text{ if } \left(\hat{x},\hat{y}\right) \in \OmegaMouth,\\  
         0 &\text{ if } \left(\hat{x},\hat{y}\right) \notin \OmegaMouth 
    \end{cases}
\end{equation}
is a piecewise constant function, where $\VelScle$ is the exhalation velocity, and $\OmegaMouth \subset \left[0,\hat{a}\right]\times\left[0,\hat{b}\right]$ is the location of the wearer's mouth on the surface $z=f(x,y)$. On the rear of the mask, we impose continuity of flux, pressure, and following the discussion in \cite{Pereira2021}, a no-slip condition,
\begin{equation}
\label{BC:mask interface}
        \hat{\uvec} \btimes \nvec_{g} =0, \quad \hat{\uvec} \bcdot \nvec_{g} = \phi \hat{\Uvec}\bcdot \nvec_{g}, \quad\hat{p}=\hat{P} \quad \text{ at } \hat{z} =  \hat{g}\left(\hat{x},\hat{y}\right).
\end{equation}
We assume that the air surrounding the mask and wearer is at atmospheric pressure. Consequently, we set the pressure in front of the mask
\begin{equation}
    \hat{P} = \hat{p}_{\atm} \quad \text{ at } \hat{z} =  \hat{g}\left(\hat{x},\hat{y}\right)+\hat{T}\sqrt{1+  \left( \frac{\partial \hat{g}}{\partial \hat{x}}\right)^2 + \left( \frac{\partial \hat{g}}{\partial \hat{y}}\right)^2}\label{eq:DimBCS_front}
\end{equation}
We also set the pressure to be atmospheric at any open boundaries of mask material. A traditional face mask will be open on all sides, so we set
\begin{equation}
    \hat{p} = \hat{p}_{\atm}, \quad \hat{P} = \hat{p}_{\atm}  \quad \text{ if } \hat{x}=0,\,\hat{a}, \text{ or } \hat{y} = 0,\,\hat{b}.\label{eq:DimBCS_sides_mask}
\end{equation}
A neck gaiter is open only on the top and bottom, so we set 
\begin{subequations}
\label{eq:DimBCS_sides_gaiter}
\begin{align}
       \hat{p} = \hat{p}_{\atm}, \quad \hat{P} = \hat{p}_{\atm}  \quad& \text{ if }  \hat{y} = 0,\,\hat{b}.
\end{align}
along with a periodic condition in the $x$-direction, 
\begin{align}
    \hat{\uvec}, ~ \hat{p}, ~ \hat{P} ~ \text{ periodic } \quad & \text{ at } \hat{x}=0,\,\hat{a}.
\end{align}
\end{subequations}
To complete the specification of the fluid problem, we assume that the fluid velocity is initially zero, \textit{i.e.} we write
\begin{equation}
    \hat{\uvec} = \boldsymbol{0}\quad \text{ at } \hat{t} = 0.\label{eq:Dim_IC_fluid}
\end{equation}

\subsection{Aerosol transport modelling}

We now consider a simple model for the transport of viral-containing aerosols by the airflow. We assume that the aerosols have negligible volume fraction and do not affect the airflow. We further assume that all the aerosols are identical, have constant physical properties, and are transported by advection alone. Thus, in the gap between the face covering and the wearer's face, their distribution is described by the transport equation 
\begin{equation}
    \frac{\partial \hat{c}}{\partial \hat{t}}+ \hat{\uvec} \bcdot \hat{\bnabla}\hat{c}  = 0.\label{eq:Dim_Gap_conc}
\end{equation}
where $\hat{c}$ is the concentration of aerosols,  measured in \unit{mol.m^{-3}}. 
In the mask, we will assume that all the pores are identical, and the number of aerosol particles is small compared with the number of pores, so by \cite{Wey2023}, the appropriate transport equation is 
\begin{equation}
    \frac{\partial \hat{C}}{\partial \hat{t}}+\phi \hat{\Uvec}\bcdot \hat{\bnabla}\hat{C} = - \frac{\hat{\gamma}\left|\hat{\Uvec} \right| \hat{C}}{\phi},\label{eq:Dim_Mask_conc}
\end{equation}
where $\hat{C}$ is the (local average) concentration of aerosols and $\hat{\gamma}\left|\hat{\Uvec} \right| $ is the rate of particle capture by the filter. We further assume for simplicity that the capture of aerosols does not result in a change in the porosity, permeability, or capture rate of the mask. Finally, we must impose boundary conditions for the aerosol. Since we are considering the airflow during exhalation, we will focus on the outwards protection offered by the face covering, and so assume that the viral aerosols are being generated by the wearer. Consequently, we impose the concentration at the mouth,
\begin{equation}
    \hat{c} = 
    \begin{cases}
        \hat{c}_{\In} & \text{ if }\left(\hat{x},\hat{y}\right) \in \OmegaMouth\\
        0 & \text{ if }\left(\hat{x},\hat{y}\right) \notin \OmegaMouth
    \end{cases},
    \quad \text{ at } \hat{z} = \hat{f}\left(\hat{x},\hat{y}\right)
    ,\label{eq:DimBCS_face_conc}
\end{equation}
and continuity in the flux of aerosols across the rear surface of the mask, 
\begin{equation}
    \hat{c}\hat{\uvec} \bcdot \nvec_{g} = \phi \hat{C}\hat{\Uvec}\bcdot \nvec_{g}, \quad  \text{ at } \hat{z} =  \hat{g}\left(\hat{x},\hat{y}\right).\label{eq:DimBCS_rear_conc}
\end{equation}
Conditions on the remaining boundaries are not necessary due to the hyperbolic nature of \eqref{eq:Dim_Gap_conc}--\eqref{eq:Dim_Mask_conc}. Finally, we assume that there are no viral aerosols in the gap or face covering initially, and so we write 
\begin{equation}
    \hat{c} = 0, \quad \hat{C} = 0 \quad \text{ at } \hat{t} = 0.\label{eq:Dim_IC_conc}
\end{equation}

\subsection{Nondimensionalisation}
Our model comprises the six equations \eqref{eq:Dim_NS_mass}--\eqref{eq:Dim_Mask_Darcy} and \eqref{eq:Dim_Gap_conc}--\eqref{eq:Dim_Mask_conc}, subject to the boundary conditions \eqref{eq:DimBCS_face}--\eqref{eq:DimBCS_front} and \eqref{eq:DimBCS_face_conc}--\eqref{eq:DimBCS_rear_conc}, along with \eqref{eq:DimBCS_sides_mask} for a face mask or \eqref{eq:DimBCS_sides_gaiter} for a neck gaiter, and initial conditions \eqref{eq:Dim_IC_fluid} and \eqref{eq:Dim_IC_conc} for the six variables $\hat{\uvec}$, $\hat{p}$, $\hat{\Uvec}$, $\hat{P}$, $\hat{c}$, and $\hat{C}$. We nondimensionalise the model using the following scalings
\begin{align}
     \left(\hat{x},\hat{y},\hat{z}\right) &=  \left(\LgthScle x,\LgthScle y, \hat{h}_{0} z \right), ~& \left( \hat{f},\hat{g}\right)&= \hat{h}_{0}\left(f,g \right), ~ &\hat{t}&=\frac{\hat{b}^2 \hat{h}_{0}}{\hat{A} \hat{w}^{*}}t,~&  \left(\hat{c},\hat{C}\right)& = \hat{c}_{\In}\left(c,C \right)\\[0.5ex]      
    \left(\hat{u},\hat{v},\hat{w}\right) &= \frac{\hat{A} \hat{w}^*}{\hat{b} \hat{h}_{0}}\left(u,v,\epsilon w \right), ~&  \left(\hat{U},\hat{V},\hat{W}\right) &= \frac{\epsilon\hat{A} \hat{w}^*}{\hat{b} \hat{h}_{0}}\left(\epsilon U,\epsilon V,W \right),~ & \hat{w}_{\In}& = \frac{\epsilon\hat{A} \hat{w}^{*}}{\hat{b}\hat{h}_0}w_{\In},~&\left(\hat{p},\hat{P}\right) & = \hat{p}_{\atm}(1,1) +  \frac{\hat{\mu} \hat{w}^* \hat{A}}{\hat{h}_{0}^3 }\left(p,P\right).
    \label{eq:Nondim_Scalings}
\end{align}

Here, $\epsilon  = \hat{h}_{0}/\hat{b}$ is the aspect ratio of the gap between the face covering and the face, where $\hat{h}_{0}$ is a typical thickness (equal to the mean of $\hat{h}(x,y)$ over the face). The velocity scale has been chosen so that there is a unit total dimensionless flux through the mouth, and we have chosen the pressure scaling to balance the pressure gradient along the gap between the face covering and the wearer's face with viscous forces in the $z$-direction. 
Applying the scalings \eqref{eq:Nondim_Scalings} to the space between the mask and the wearer's face, \eqref{eq:Dim_NS_mass}--\eqref{eq:Dim_NS_momentum} and \eqref{eq:Dim_Gap_conc}, gives
\begin{align}
    \bnabla \bcdot \uvec & = 0,\label{eq:NondimEQNS_gap_mass}\\
    \epsilon^2 \Rey \left( \frac{\partial u}{\partial t}+ \uvec \bcdot \bnabla u \right) & = - \frac{\partial p}{\partial x} + \epsilon^2 \left(\frac{\partial^2 u}{\partial x^2} + \frac{\partial^2 u}{\partial y^2} \right) + \frac{\partial^2 u}{\partial z^2},\label{eq:NondimEQNS_gap_NSx}\\
    \epsilon^2 \Rey \left( \frac{\partial v}{\partial t}+ \uvec \bcdot \bnabla v \right) & = - \frac{\partial p}{\partial y} + \epsilon^2 \left(\frac{\partial^2 v}{\partial x^2} + \frac{\partial^2 v}{\partial y^2} \right) + \frac{\partial^2 v}{\partial z^2},\label{eq:NondimEQNS_gap_NSy}\\
    \epsilon^2 \Rey \left(\frac{\partial w}{\partial t}+ \uvec \bcdot \bnabla w \right) & = - \frac{1}{\epsilon^2 }\frac{\partial p}{\partial z} + \epsilon^2 \left(   \frac{\partial^2 w}{\partial x^2}  + \frac{\partial^2 w}{\partial y^2} \right) + \frac{\partial^2 w}{\partial z^2},\label{eq:NondimEQNS_gap_NSz}\\
    \frac{\partial c}{\partial t}+  \uvec \bcdot \bnabla c   &=0,\label{eq:NondimEQNS_gap_conc}
\end{align}
where $\Rey = \hat{\rho}\hat{A} \hat{w}^*/\left(\hat{\mu} \hat{h}_{0} \right) $ is the Reynolds number. The dimensionless governing equations within the mask, \eqref{eq:Dim_Mask_mass}--\eqref{eq:Dim_Mask_Darcy} and \eqref{eq:Dim_Mask_conc}, are
\begin{align}
    \epsilon^2\left( \frac{\partial}{\partial x}\left( \phi U \right) + \frac{\partial }{\partial y} \left( \phi V \right) \right) + \frac{\partial}{\partial z}\left( \phi W \right) & = 0,\label{eq:NondimEQNS_mask_mass}\\
     \phi U &= -K \frac{\partial P}{\partial x},\label{eq:NondimEQNS_mask_Darcyx}\\
    \phi V &= -  K \frac{\partial P}{\partial y},\label{eq:NondimEQNS_mask_Darcyy}\\
    \phi W &= - K \frac{\partial P}{\partial z},\label{eq:NondimEQNS_mask_Darcyz}\\
    \frac{\partial C}{\partial t} + \epsilon^2\left( \phi U\frac{\partial C}{\partial x} + \phi V\frac{\partial C}{\partial y} \right) + \phi W\frac{\partial C}{\partial z}  &= - \Gamma \left( \epsilon^2 \left(U^2 +  V^2  \right) + \left(W\right)^2 \right)^{1/2} C,
    \label{eq:NondimEQNS_mask_conc}
\end{align}
where 
\begin{equation}
    K = \frac{\hat{K}}{\epsilon^4 \LgthScle ^2} ,\quad \Gamma = \frac{\hat{\gamma} \hat{b} }{\phi}\label{eq:3d_def_Da_Gamma}
\end{equation}
are the dimensionless permeability and the dimensionless capture number, respectively.
The boundary conditions, \eqref{eq:DimBCS_face}--\eqref{eq:DimBCS_front}  and \eqref{eq:DimBCS_face_conc}--\eqref{eq:DimBCS_rear_conc}, become
\begin{equation}
    \uvec \btimes \nvec_{f} = \boldsymbol{0}, \quad \uvec \bcdot \nvec_{f} =w_{\In}(x,y),\quad  c = \begin{cases}
        1 & \text { if } (x,y) \in\OmegaMouth\\
        0 & \text{ if } (x,y)\notin\OmegaMouth
    \end{cases} ~ \text{ at } z=f(x,y),\label{eq:NondimBC_face}
\end{equation}
    \begin{equation}
        \uvec \btimes \nvec_{g} = \boldsymbol{0},\quad - \left(u \frac{\partial g}{\partial x} + v \frac{\partial g}{\partial y}\right) + w 
 = - \epsilon^2 \left( \phi U \frac{\partial g}{\partial x} + \phi V \frac{\partial g}{\partial y}\right) + \phi W ,\quad   p = P , ~ c=C
     ~ \text{ at } z= g(x,y),\label{eq:NondimBC_mask_rear}
\end{equation}
     \begin{equation}
    P = 0 ~ \text{ at } z= g(x,y)+T\sqrt{1+\epsilon^2 \left( \frac{\partial g}{\partial x}\right)^2 + \epsilon^2\left( \frac{\partial g}{\partial y}\right)^2},
    \label{eq:NondimBC_mask_front}
\end{equation}
where
\begin{equation}
    w_{\In}\left(x,y\right) = 
    \begin{cases}
         \dfrac{1}{M} & \text{ if }\left(x,y\right) \in \OmegaMouth\\  
         0 &\text{ if } \left(x,y\right) \notin \OmegaMouth
    \end{cases},
\end{equation}
where $M=\hat{A}/\hat{b}^2$ is the dimensionless area of the mouth and $T=\hat{T}/\hat{h}_{0}$ is the ratio of the mask width to the reference gap width. (As mentioned previously, our scalings result in $\iint_{\OmegaMouth}w_{\In}\,\rm{d}\OmegaMouth=1$, since $\OmegaMouth$ has dimensionless area $M$).
The conditions specific to a face mask, \eqref{eq:DimBCS_front}, become 
\begin{equation}
    p=0,~P=0 ~ \text{ if } x=0,l, \text{ or } y= 0,1,\label{eq:NondimBC_sides_mask}
\end{equation}
where $l=\hat{a}/\hat{b}$ is the lateral aspect ratio and those for a neck gaiter, \eqref{eq:DimBCS_sides_gaiter}, become
\begin{subequations}
\label{eq:NondimBC_sides_gaiter}
\begin{align}
    p=0,~P=0 ~ &\text{ if } y= 0,1, \\
    \uvec,~p,~P~\text{periodic} ~& \text{ at } x= 0, l.\label{eqn:breward1}
\end{align}
\end{subequations}

The model \eqref{eq:NondimEQNS_mask_mass}--\eqref{eq:NondimBC_sides_gaiter} depends on seven dimensionless parameters: the aspect ratio $\epsilon$, the reduced Reynolds number $\epsilon^2 \Rey$, the dimensionless permeability $K$,
the dimensionless capture number $\Gamma$,
the ratio of the mask area to the mouth area $M$,  the dimensionless thickness of the mask $T$, and the mask's lateral aspect ratio $l$. In addition, the definitions of $f(x,y)$, $g(x,y)$, and $\OmegaMouth$ will contain dimensionless geometric factors. In Table \ref{tab:dimensional}, we present the range of typical values of the dimensional parameters, representing a wearer at rest. With the exception of mean gap width, we will use fixed reference values to estimate the value of the dimensionless parameters, given in  Table~\ref{tab:dimensionless}. Since variation in mean gap width can have a significant impact on the behaviour of our model, we will consider how the behaviour changes over the range of gap widths from an excellently fitting to a badly fitting face mask, based on mean gap sizes reported from simulations in \cite{Ni2023}. The reference value for $\hat{h}_{0}$ corresponds to the mean gap width reported for a typical fit with a gap adjacent to the nose in  \cite{Ni2023}. Since no corresponding simulations exist for a neck gaiter, we will use these values for both designs of face covering. No data is currently available on the virus denature constant, $\hat{\gamma}$. 

From Table~\ref{tab:dimensionless}, we see that $\epsilon$ and $\epsilon^2\Rey$ are small and so we will work henceforth in the limit where $\epsilon\ll1$ and $\epsilon^2\Rey\ll1$, the standard approximations that lead to lubrication flows \cite{OckendonViscous}. For the other parameters (including the dimensionless virus denature constant), we assume that they are $\textit{O}(1)$, although we note that $M$ is small (but not as small as $O(\epsilon^2)$). There are geometric parameters contained in the descriptions of the face and face covering. We will assume that variations in $f$ and $g$ are long wavelength, so that gradients are at most $O(1)$.

\begin{table}
\caption{Typical values of dimensional parameter values in the problem. We will use the reference values for all parameters except gap width for the remainder of the paper. Separate values are given for the permeabilities of different face covering materials. }
\label{tab:dimensional}
\begin{ruledtabular}
\begin{tabular}{lcccr}
Parameter & & Range & Reference Value & Source  \\
\hline
Mask width & $\hat{a}$ && \SI{1.8e-1}{m} &\\
Neck Gaiter width & $\hat{a}$ &&  \SI{5.4e-1}{m}  &  \footnotemark[1] \\
Mask height & $\hat{b}$ && \SI{9e-2}{m}&\\
Mask thickness & $\hat{T}$ && \SI{4e-4}{m}   &\cite{Bourrianne2021} \\
Mouth area & $\hat{A}$& \SIrange{1.2e-4}{1.8e-4}{m^2} &\SI{1.2e-4}{m^{2}} & \cite{Abkarian2020Speech}\\
Mean gap thickness & $\hat{h}_{0}$ & \SIrange{3.6e-4}{1.1e-3}{m} & \SI{4.9e-4}{m}&\cite{Ni2023}\footnotemark[2]\\
Density of air  & $\hat{\rho}$ & &\SI{1.2}{kg.m^{-3}}&\cite{Ni2023}\footnotemark[3]\\
Viscosity of air & $\hat{\mu}$&& \SI{1.8e-5}{Pa.s}&\cite{Ni2023}\footnotemark[3]\\
Exhalation velocity & $\VelScle$ & \SIrange{0.3}{0.7}{m.s^{-1}} & \SI{0.5}{m.s^{-1}} & \cite{Viola2021face} \\
Mask permeability (surgical mask) & $\hat{K}$ &  \SIrange{9.9e-12}{1.4e-11}{m^2} & \SI{1.0e-11}{m^2}  & \cite{Ni2023}\\
Mask permeability (neck gaiter) & $\hat{K}$ &  \SIrange{2.4e-11}{5.2e-11}{m^2} &\SI{5.0e-11}{m^2}& Supplied by Virustatic\\
Virus denature constant & $\hat{\gamma}$ && \textrm{Unknown} &
\end{tabular}
\end{ruledtabular}
\footnotetext[1]{Whilst worn, due to stretching}\\
\footnotetext[2]{Mean perimeter gap thickness for face masks with different fits predicted by simulations. }
\footnotetext[3]{At \SI{20}{\degreeCelsius}.}

\end{table}

\begin{table}
\caption{Values of the dimensionless parameter values in the model based on reference values for the dimensional parameters in Table \ref{tab:dimensional} and the given dimensional range of $\hat{h}$.}
\label{tab:dimensionless}
\begin{ruledtabular}
\begin{tabular}{ccc}
Dimensionless parameter & Formula & Value \\
\hline
\\[0ex]
$\epsilon$& $\dfrac{\hat{h}_{0}}{\hat{b}}$ & \numrange{4e-3}{1.2e-2} \\[3ex]
$l$ (surgical mask) & $\dfrac{\hat{a}}{\hat{b}}$ &\num{2.0} \\[2ex]
$l$ (neck gaiter) & $\dfrac{\hat{a}}{\hat{b}}$ & \num{6.0} \\[2ex]
$T$ & $\dfrac{\hat{T}}{\hat{h}_{0}}$ & \numrange{0.36}{1.1}  \\[2ex]
$\epsilon^2 \Rey $& $\dfrac{\hat{\rho} \hat{A} \hat{w}^{*}\hat{h}_{0}}{\hat{b}^2 \hat{\mu}}$&\numrange{0.18}{0.54} \\[2ex]
$\Gamma$& $\dfrac{\hat{\gamma} \hat{b}}{\phi} $ &Unknown \\[2ex]
$M$ & $\dfrac{\hat{A}}{\hat{b}^2}$ & \num{1.4e-2}\\[2ex]
$K$ (surgical mask) &$\dfrac{\hat{K}\hat{b}^2}{\hat{h}^4}$ & \numrange{5.5e-2}{4.8} \\[2ex]
$K$ (neck gaiter) &$\dfrac{\hat{K}\hat{b}^2}{\hat{h}^4}$ & \numrange{2.8e-1}{2.4e1} \\[3ex]
\end{tabular}
\end{ruledtabular}
\end{table}

\section{Leading-order flow model}\label{sec:Asymptotics}

We now seek a  steady solution of \eqref{eq:NondimEQNS_gap_mass}--\eqref{eq:NondimBC_sides_gaiter} in the limit of small $\epsilon$ and $\epsilon^2\Rey$, and we neglect terms of these sizes henceforth. As we only consider the leading-order behaviour, we will not introduce extra notation to distinguish this. We will first solve for the flow through the mask and then for the flow in the gap between the face and face covering.

\subsection{Flow through the mask}
The leading-order versions of \eqref{eq:NondimEQNS_mask_mass}--\eqref{eq:NondimEQNS_mask_Darcyz} for flow in the mask  are 
\begin{align}
    \frac{\partial}{\partial z}\left( \phi W \right) = 0,\quad
    \phi U = - K \frac{\partial P}{\partial x},\quad
    \phi V = - K \frac{\partial P}{\partial y},\quad 
    \phi W = - K \frac{\partial P}{\partial z},\label{eq:Asympt_EQN_mask_Darcy}
\end{align}
for $g(x,y)<z<g(x,y)+T$, and are subject to the leading order of the boundary condition at the front of the mask, \eqref{eq:NondimBC_mask_front}, 
\begin{equation}
    P = 0 \quad \text{ at } z=g(x,y)+T.\label{eq:Asympt_BC_mask_front}
\end{equation}
We solve the first and last equation in \eqref{eq:Asympt_EQN_mask_Darcy}, using \eqref{eq:Asympt_BC_mask_front}, to find
\begin{equation}
    \phi W = Q(x,y), \quad P(x,y) = \frac{Q(x,y)}{K }\left( g(x,y) + T - z \right),\label{eq:Asympt_Mask_SOL}
\end{equation}
where $Q(x,y)$ is the unknown (Darcy) flux of air through the mask at each point. We note that the remaining equations in \eqref{eq:Asympt_EQN_mask_Darcy}, provide expressions for $U$ and $V$ in terms of $Q$.  
\subsection{Flow in the gap between face and face covering}
The leading-order versions of the equations for the flow in the gap between the face and the face covering \eqref{eq:NondimEQNS_gap_mass}--\eqref{eq:NondimEQNS_gap_NSz} are
 \begin{align}
    \frac{\partial u}{\partial x} + \frac{\partial v}{\partial y} + \frac{\partial w}{\partial z}  & = 0,\label{eq:Asympt_EQN_channel_mass}\\
    0 & = - \frac{\partial p}{\partial x} + \frac{\partial^2 u}{\partial z^2},\label{eq:Asympt_EQN_channel_mommentum_x}\\
    0 & = - \frac{\partial p}{\partial y} + \frac{\partial^2 v}{\partial z^2},\label{eq:Asympt_EQN_channel_mommentum_y}\\
    0 & = - \frac{\partial p}{\partial z},\label{eq:Asympt_EQN_channel_mommentum_z}
\end{align}
for $ f(x,y)<z<g(x,y)$, and, using the solution for the mask, the leading-order dimensionless boundary conditions \eqref{eq:NondimBC_face}--\eqref{eq:NondimBC_mask_rear}  are
\begin{align}
    u= 0, ~ v= 0, ~ w= w_{\In}(x,y),~ \text{ at } z= f(x,y),\label{eq:Asympt_BC_face}\\
    u= 0, ~ v= 0, ~ w = Q(x,y), ~ p =\frac{T}{K }Q(x,y),~ \text{ at } z= g(x,y),\label{eq:Asympt_BC_rear}
\end{align}
The leading-order boundary conditions on the perimeter of the mask are 
\begin{equation}
    p = 0, ~ \text{ if } x=0,l \text{ or } y=0,1,\label{eq:Asympt_BC_sides_mask}
\end{equation}
for a face mask,  using \eqref{eq:NondimBC_sides_mask}, and 
\begin{subequations}
    \label{eq:Asympt_BC_sides_gaiter}
    \begin{align}
            p = 0, ~ \text{ if }  y=0,1,\\
        p ~ \text{ periodic } ~\text{ at } x= 0,l,
    \end{align}
    for a neck gaiter, using \eqref{eq:NondimBC_sides_gaiter}. 
\end{subequations}

\begin{figure}[h]
    \centering
    \includegraphics[width=0.4\textwidth]{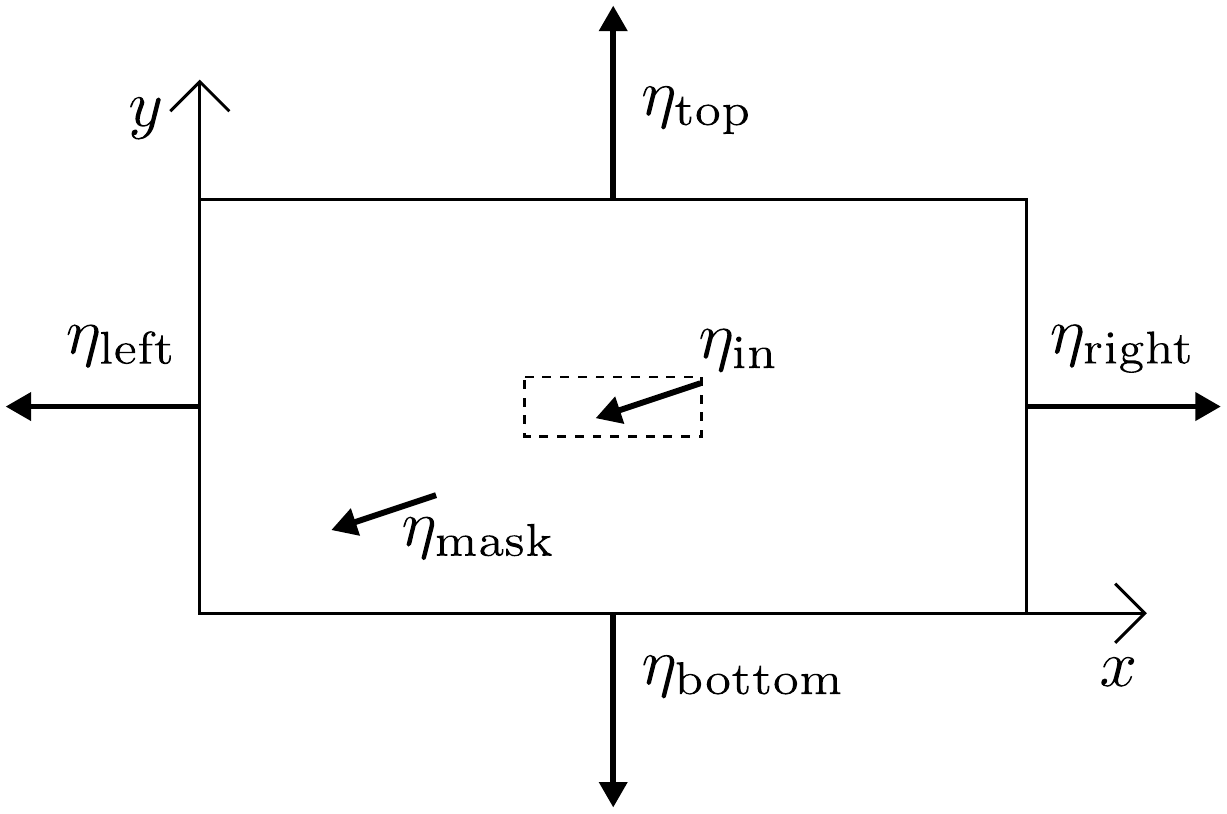}
    \caption{The fluxes through each boundary of a face covering defined in \eqref{eq:Def_totalflux}. The dashed rectangle indicates the position of the mouth below the mask, which is indicated by the solid rectangle.}
    \label{fig:flux_diagram}
\end{figure}

Equation
\eqref{eq:Asympt_EQN_channel_mommentum_z} tells us that the pressure in the gap does not vary with $z$ and so, using \eqref{eq:Asympt_BC_rear}, we find that
\begin{equation}
    p= 
    \frac{T}{K }Q(x,y).\label{eq:Asympt_SOL_pressure}
\end{equation}
Consequently, we integrate \eqref{eq:Asympt_EQN_channel_mommentum_x} and \eqref{eq:Asympt_EQN_channel_mommentum_y} and apply \eqref{eq:Asympt_BC_face} and \eqref{eq:Asympt_BC_rear} to find
\begin{subequations}
\label{eq:Asympt_Sol_velocity}
\begin{align}
    u &= \frac{1}{2}\frac{\partial p}{\partial x}\left( z- f \right)\left( z - g \right), \\ v &= \frac{1}{2}\frac{\partial p}{\partial y}\left( z- f \right)\left( z - g \right).
\end{align}
\end{subequations}
We thus calculate the total flux parallel to the face in the gap between the face and the face covering, $\qvec = \left(q_{x},q_{y}\right)$, to be given by, in component form,
\begin{subequations}
\label{eq:Asympt_sol_q}
\begin{align}
    q_{x}(x,y) =\int_{f(x,y)}^{g(x,y)} u \, \mathrm{d}z = -\frac{h^3}{12}\frac{\partial p}{\partial x},\\
    q_{y}(x,y) =\int_{f(x,y)}^{g(x,y)} v \, \mathrm{d}z =  -\frac{h^3}{12}\frac{\partial p}{\partial y},
\end{align}
\end{subequations}
where $h(x,y) = g(x,y) - f(x,y)$ is the (dimensionless) gap width. Integrating the conservation of mass equation \eqref{eq:Asympt_EQN_channel_mass} and applying the flux boundary conditions \eqref{eq:Asympt_BC_face} and \eqref{eq:Asympt_BC_rear}, yields the relationship
\begin{equation}
    \frac{\partial q_{x}}{\partial x} + \frac{\partial q_{y}}{\partial y}  + Q = w_{\In}.\label{eq:Asympt_mass_integral}
\end{equation}
Substituting \eqref{eq:Asympt_SOL_pressure} into \eqref{eq:Asympt_sol_q} and combining with \eqref{eq:Asympt_mass_integral}, we arrive at a single partial differential equation for the flux $Q$, namely
\begin{equation}
    \nabla \cdot \left( h^3 \nabla Q \right) - k^2 Q = - k^2 w_{\In},\label{eq:Asympt_Reduced_gov_eqn}
\end{equation}
where $k^2=12K/T $. Equation \eqref{eq:Asympt_Reduced_gov_eqn}  is an inhomogeneous modified Helmholtz equation \cite{OckendonPDEs}. 
Using \eqref{eq:Asympt_BC_sides_mask} and \eqref{eq:Asympt_SOL_pressure}, we see that appropriate boundary conditions for $Q$ for a face mask are
\begin{equation}
    Q = 0 \text{ at } x=0,~ x=l,~ y=0, ~y=1,\label{eq:Asympt_Reduced_bc}
\end{equation}
while \eqref{eq:Asympt_BC_sides_gaiter} and \eqref{eq:Asympt_SOL_pressure} give the appropriate conditions for a neck gaiter,
\begin{subequations}
    \label{eq:Asympt_Reduced_bc_periodic}
    \begin{align}
    Q = 0 &\text{ at } y=0,~ y=1,\\
        Q ~ \text{ periodic } &\text{ at } x=0,~ x=l.\label{eq:Asympt_Reduced_bc_periodic_Periodic}
    \end{align}
\end{subequations}
We therefore have a closed problem \eqref{eq:Asympt_Reduced_gov_eqn}--\eqref{eq:Asympt_Reduced_bc_periodic} for $Q$, and using its solution, may subsequently determine $W$, $P$, $p$, $w$, $u$, and $v$ by using, \eqref{eq:Asympt_Mask_SOL}, \eqref{eq:Asympt_EQN_channel_mass}, \eqref{eq:Asympt_SOL_pressure}, and \eqref{eq:Asympt_sol_q}. Given a solution to \eqref{eq:Asympt_Reduced_gov_eqn}--\eqref{eq:Asympt_Reduced_bc_periodic}, we calculate the total flux through each boundary in the problem. We therefore define
\begin{subequations}
\label{eq:Def_totalflux}
\begin{align}
    \eta_{\textrm{mask}} &= \int_{0}^{1}\int_{0}^{l}Q(x,y)\,\mathrm{d}x\mathrm{d}y, &
    \eta_{\textrm{in}}&= \int_{0}^{1}\int_{0}^{l}w_{\In}(x,y)\,\mathrm{d}x\mathrm{d}y,
        \\
    \eta_{\textrm{left}} &= -\int_{0}^{1} q_{x}(0,y) \,\mathrm{d}y, &
    \eta_{\textrm{right}} &= \int_{0}^{1} q_{x}(l,y) \,\mathrm{d}y,\\
    \eta_{\textrm{bottom}} &= - \int_{0}^{l} q_{y}(x,0) \,\mathrm{d}x, & 
    \eta_{\textrm{top}} &= \int_{0}^{l} q_{y}( x,1)\mathrm{d}x,  
\end{align}
\end{subequations}
as the magnitude of the total flux through the mask, out of the mouth, and through the left, right, bottom and top boundaries, respectively. By our choice of scaling we know that $\eta_{\In}=1$, implying that $\eta_{\textrm{mask}}$, $\eta_{\textrm{left}}$, $\eta_{\textrm{right}}$, $\eta_{\textrm{top}}$, and $\eta_{\textrm{bottom}}$, are just the proportion of the total flux out their boundaries, respectively, as shown in Fig.~\ref{fig:flux_diagram}.
It follows that the leakage ratio, $\eta$, is given by
\begin{equation}
    \eta = \frac{\eta_{\textrm{left}}+\eta_{\textrm{right}}+\eta_{\textrm{bottom}}+\eta_{\textrm{top}}}{\eta_{\textrm{in}}} =\eta_{\textrm{left}}+\eta_{\textrm{right}}+\eta_{\textrm{bottom}}+\eta_{\textrm{top}}.\label{eq:Def_eta}
\end{equation}
Often, we will consider configurations with horizontal or vertical symmetry, and therefore define
\begin{equation}
    \eta_{\hoz } = \frac{\eta_{\textrm{left}}+\eta_{\textrm{right}}}{\eta_{\textrm{in}}}, \quad  \eta_{\ver  }= \frac{\eta_{\textrm{top}}+\eta_{\textrm{bottom}}}{\eta_{\textrm{in}}},\label{eq:Def_eta_hoz_vert}
\end{equation}
as the total proportion of the flux leaking out the horizontal and vertical boundaries, respectively (and note that $\eta=\eta_{\hoz }+\eta_{\ver }$).  
 Integrating  \eqref{eq:Asympt_Reduced_gov_eqn} over the mask implies $ 1=\eta_{\textrm{mask}}+\eta_{\textrm{left}}+\eta_{\textrm{right}}+\eta_{\textrm{bottom}}+\eta_{\textrm{top}}$, which expresses the total conservation of mass, and provides the alternate expression $\eta = 1- \eta_{\textrm{mask}}$.

\section{Results for leakage}\label{sec:Leakage}

\subsection{One-dimensional results}

We briefly consider the solution to our reduced model for a face mask in one dimension to gain intuition about its behaviour. For simplicity, we consider only the case $l=2$, with the mouth, of width $2\mathcal{M}$, centred on $x=1$, allowing us to use symmetry to only consider half of the domain. We also assume the gap width $h$ is constant ($h=1$) so that the problem is linear, and \eqref{eq:Asympt_Reduced_gov_eqn}--\eqref{eq:Asympt_Reduced_bc} reduce to
\begin{equation}
    \frac{\mathrm{d}^2 Q}{\mathrm{d}x^2}  -  k ^2 Q = -  k ^2 w_{\In}(x),\label{eq:1D_Reduced_GOVEQN}
\end{equation}
where 
\begin{equation}
   w_{\In}(x) = \begin{cases}
        0 & \text{ if } 0<x<1-\mathcal{M}\\[0.5ex]
        1/\mathcal{M} & \text{ if } 1-\mathcal{M}<x<1
    \end{cases},
\end{equation}
subject to
\begin{equation}
    Q = 0 \text{ at } x=0,\quad \frac{\mathrm{d}Q}{\mathrm{d}x} =0 \text{ at } x= 1.\label{eq:1D_Reduced_BC}
\end{equation}
By requiring continuity of $Q$ and its derivative at $x=1-\mathcal{M}$, we solve \eqref{eq:1D_Reduced_GOVEQN}--\eqref{eq:1D_Reduced_BC} to find
\begin{equation}
    Q(x) = 
    \begin{cases}
        \displaystyle\frac{\sinh\left(  k  \mathcal{M} \right) \sinh\left(  k  x \right)}{\mathcal{M}\cosh\left( k  \right)} & \text{ if } 0<x<1-\mathcal{M}\\[0.5ex]
        \displaystyle\frac{1}{\mathcal{M}}- \displaystyle\frac{\cosh\left(  k \left(1-\mathcal{M} \right) \right)\cosh\left(  k  \left( x-1 \right)\right)}{\mathcal{M}\cosh\left( k  \right)} & \text{ if } 1-\mathcal{M}<x<1.
    \end{cases}
\end{equation}
From this, we use conservation of mass to calculate a closed-form expression for the leakage ratio,
\begin{equation}
    \eta = 1- \int_{0}^{1}Q(x)\,\mathrm{d}x =  \frac{\sinh\left(\mathcal{M}  k  \right)}{ k  \mathcal{M} \cosh\left( k  \right)}.\label{eq:1D_eta}
\end{equation}

\begin{figure}[h]
    \centering
    \includegraphics[width=0.55\linewidth]{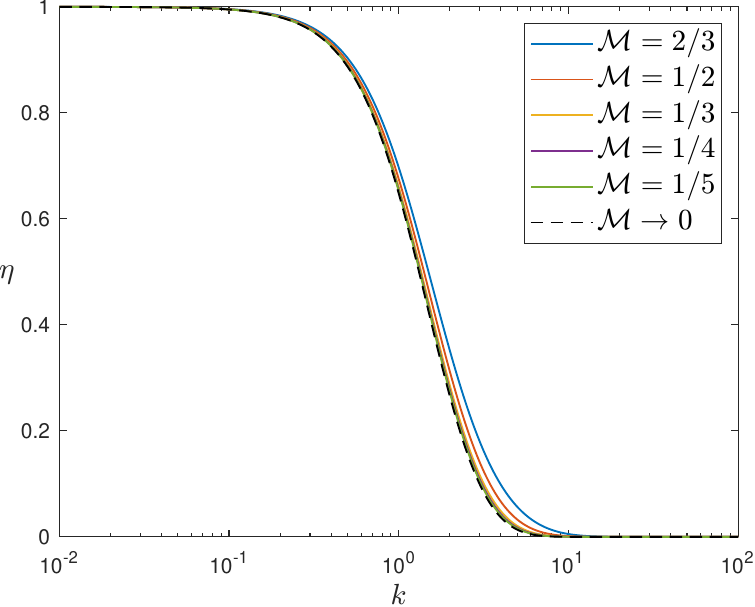}
    \caption{The leakage ratio $\eta$ as a function of $ k $ predicted by the one-dimensional reduced model for different values of $\mathcal{M}$, the ratio between the width of the mouth to the width of the mask. The limit $\mathcal{M}\to0$ is shown by a black dashed line.}
    \label{fig:1D_LeakageCurve}
\end{figure}

In Fig.~\ref{fig:1D_LeakageCurve} we plot $\eta$ as a function of $ k $ for different values of $\mathcal{M}$. We see that varying $\mathcal{M}$ only has a small effect on the amount of leakage, and, once $\mathcal{M}<1/3$ (which corresponds to a mask more than three times the width of the mouth), the leakage
is less than 2\% from the value as $\mathcal{M}\to 0 $, given by $\eta \sim 1/\cosh( k )$, and shown in the figure by a dashed line. 

Examining the variation with $ k $, we see that $\eta\to1$ as $ k  \to 0$ and $\eta \to0$ as $ k  \to \infty$. From \eqref{eq:1D_eta} it is clear that the transition between these two limits happens exponentially when $ k  \approx 1$. In terms of the dimensional parameters, $ k ^2 = 12 \LgthScle^2 \hat{K}/\left( \hat{h}_{0}^3 \hat{T}\right)$, which is the ratio between the resistance to flow through the porous material $\hat{K}\hat{b}^2/\hat{T}\hat{\mu}$, and the resistance to fully developed viscous flow through a two-dimensional channel $\hat{h}_{0}^3/12\hat{\mu}$. Therefore, small $ k $ corresponds to a low permeability and large gap width, while large $ k $ corresponds to high permeability and low gap width. A key observation is that $ k ^2$ scales linearly with the permeability and the mask thickness, but with the inverse cube of the gap width. This means that the leakage ratio will be more sensitive to changes in fit than to changes in material. Although this one-dimensional solution provides an explicit approximation for the leakage, we cannot use it to investigate the differences between a mask and a neck gaiter, as we cannot simultaneously have two ends joined to form a closed loop, and still have the possibility of leakage. We therefore now investigate the full two-dimensional version of \eqref{eq:Asympt_Reduced_gov_eqn}--\eqref{eq:Asympt_Reduced_bc_periodic}.

\subsection{Two-dimensional results: constant gap width}\label{sec:Leakage_2D_constant}

We again consider the solution under the restriction that $h$ is constant, {\it i.e.} $h\equiv1$. We begin by seeking a solution to \eqref{eq:Asympt_Reduced_gov_eqn} for a traditional face covering, so that \eqref{eq:Asympt_Reduced_bc} holds on all four boundaries. It is then clear from the boundary conditions that the appropriate series expansion is
\begin{equation}
    Q(x,y) = \sum_{m=1}^{\infty}\sum_{n=1}^{\infty} \alpha_{mn}\sin\left( \frac{m\pi x}{l}\right) \sin\left( n\pi y\right),\label{eq:2D_constant_series_Q}
\end{equation}
with the coefficients given by 
\begin{equation}
    \alpha_{mn} = \int_{0}^{1}\int_{0}^{a} Q(x,y)\sin\left( \frac{m\pi x}{l}\right)\sin\left(  n\pi y\right)\,\mathrm{d}x\mathrm{d}y.
\end{equation}
We also represent the velocity through the mouth in the same basis, and so write
\begin{equation}
    w_{\In}(x,y) = \sum_{m=1}^{\infty}\sum_{n=1}^{\infty} \beta_{mn}\sin\left( \frac{m\pi x}{l}\right) \sin\left( n\pi y\right),
\end{equation}
where
\begin{equation}
    \beta_{mn} = \frac{4}{lM}\iint_{\OmegaMouth}\sin\left( \frac{m\pi x}{l}\right)\sin\left( n\pi y\right) \,\mathrm{d}V,\label{eq:2D_constant_series_beta_general}
\end{equation}
which depends on the given shape of the mouth.
By multiplying \eqref{eq:Asympt_Reduced_gov_eqn} by the $mn$-th term in \eqref{eq:2D_constant_series_Q}, integrating by parts and using orthogonality, it is straightforward to show that $\alpha_{mn}$ satisfies
\begin{equation}
    \alpha_{mn} = \frac{ k ^2 \beta_{mn}}{\frac{m^2 \pi^2}{l^2}+n^2 \pi^2+ k ^2}.
\end{equation}
We note that $\beta_{mn}$ is independent of $ k $, so only needs computing once per mouth. For the simplest case of a rectangular mouth with centre $\left(E_{x},E_{y}\right)$, and dimensions $2L_{x}$ and $2L_{y}$ in the $x$- and $y$-directions respectively, equation \eqref{eq:2D_constant_series_beta_general} yields
\begin{equation}
    \beta_{mn} = \frac{16}{mn\pi^2M}\sin \bigg(  \frac{m\pi E_{x}}{l}\bigg)\sin \bigg(  \frac{m\pi L_{x}}{l}\bigg)
    \sin \bigg(  n\pi E_{y}\bigg)\sin \bigg(  n\pi L_{y}\bigg).\label{eq:2D_constant_series_beta_rectangle}
\end{equation}
Given \eqref{eq:2D_constant_series_Q}, we  calculate the depth-integrated flux of the air behind the mask  via \eqref{eq:Asympt_SOL_pressure} and \eqref{eq:Asympt_sol_q}, to find 
\begin{subequations}
\label{eq:2D_constant_series_q}
\begin{align}
q_{x}(x,y)  &= - \sum_{m=1}^{\infty}\sum_{n=1}^{\infty} \frac{m\pi}{l k ^2}\alpha_{mn}\cos\left(\frac{m\pi x}{l}\right) \sin\left( n\pi y\right),\\
    q_{y}(x,y)  &= - \sum_{m=1}^{\infty}\sum_{n=1}^{\infty} \frac{n\pi}{ k ^2}\alpha_{mn}\sin\left(\frac{m\pi x}{l}\right) \cos\left( n\pi y\right),
\end{align}
\end{subequations}
which are straightforwardly shown to be convergent. By evaluating \eqref{eq:2D_constant_series_q} along the boundary we calculate the flux of air leaking out of the sides at each point. We also substitute \eqref{eq:2D_constant_series_Q} and \eqref{eq:2D_constant_series_q}  into \eqref{eq:Def_totalflux} to find expressions for the total fluxes through each boundary, to find that
\begin{samepage}
\begin{subequations}
\begin{align}
    \eta_{\textrm{left}} & =  \sum_{m=1}^{\infty}\sum_{n=1}^{\infty}\frac{m}{ln k ^2} \alpha_{mn} \left( 1 - \left( -1 \right)^{n}\right),\quad&
    \eta_{\textrm{right}} & = - \sum_{m=1}^{\infty}\sum_{n=1}^{\infty} \frac{m}{ln k ^2} \alpha_{mn} \left(-1\right)^{m}\left( 1 - \left( -1 \right)^{n}\right),\\
    \eta_{\textrm{bottom}} &=    \sum_{m=1}^{\infty}\sum_{n=1}^{\infty} \frac{ln}{m k ^2} \alpha_{mn} \left( 1 - \left( -1 \right)^{m}\right),\quad&
    \eta_{\textrm{top}} & = -  \sum_{m=1}^{\infty}\sum_{n=1}^{\infty} \frac{ln}{m k ^2} \alpha_{mn} \left( 1 - \left( -1 \right)^{m}\right)\left(-1\right)^{n},
\end{align}
\vspace{-0.5cm}
\begin{equation}
    \eta_{\textrm{mask}}   = \sum_{m=1}^{\infty}\sum_{n=1}^{\infty} \frac{l}{mn\pi^2}\alpha_{mn}\left( 1 - \left( -1 \right)^{m}\right)\left( 1 - \left( -1 \right)^{n}\right), 
\end{equation}
\end{subequations}
\end{samepage}
allowing us to evaluate the leakage ratio.

\begin{figure}
    \centering
    \includegraphics[width=0.8\linewidth]{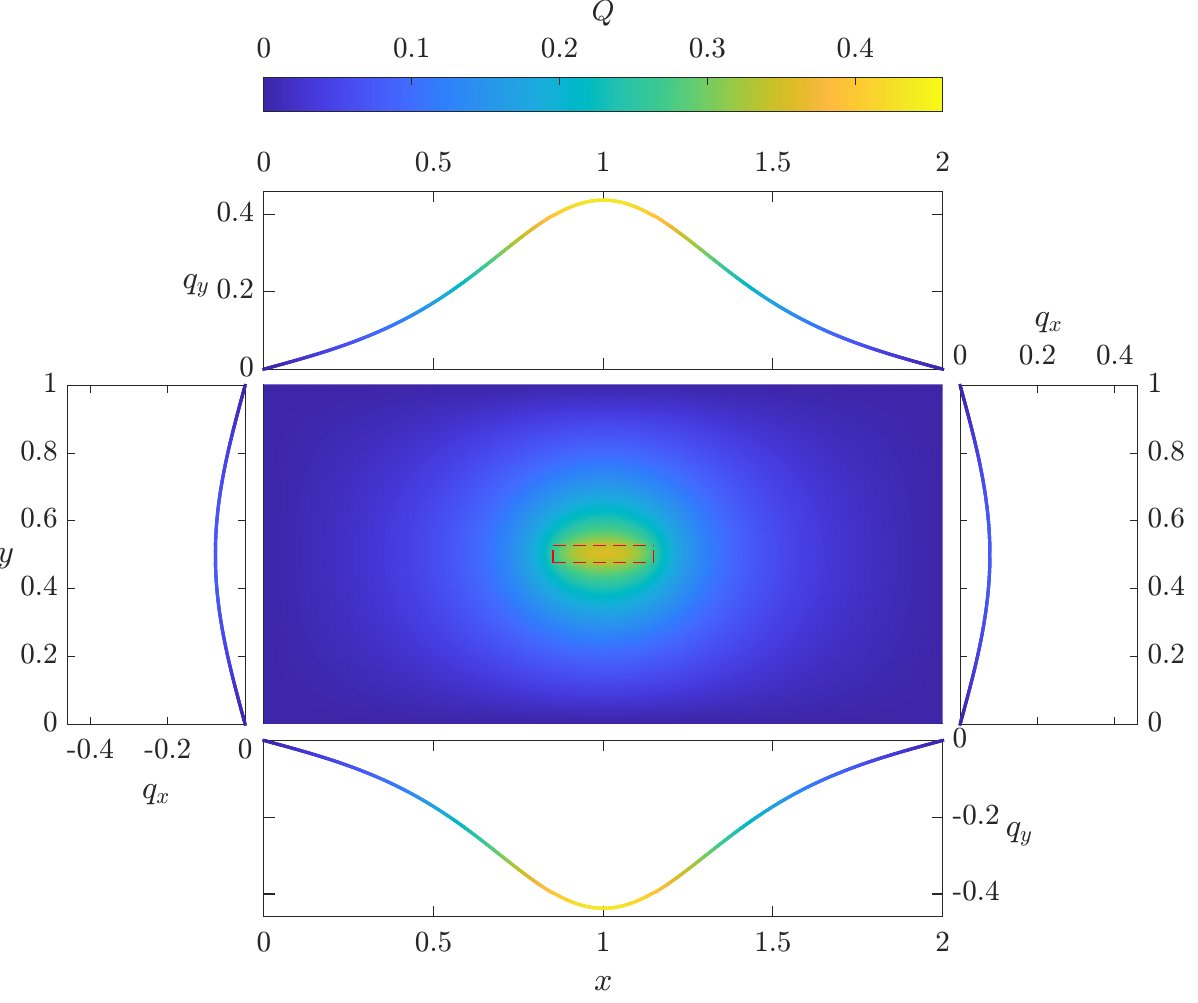}
    \caption{The central panel shows the flux $Q$ through a face mask when $ k =1$. The airflow originates from  a mouth occupying $0.852 \le x \le 1.148$, $0.475 \le y \le 0.525$, shown as a dashed red line. The corresponding fluxes arising from leakage out of the side of the mask are shown in the surrounding panels. Colour is used to show the magnitude of the flux in all panels.  }
    \label{fig:2D_Example_Solution_Mask}
\end{figure}

In Fig.~\ref{fig:2D_Example_Solution_Mask}, we show an example solution for the flux of exhaled air from a face mask with constant gap width and $ k =1$, found using \eqref{eq:2D_constant_series_Q} and \eqref{eq:2D_constant_series_q}. For numerical purposes, the infinite sums are truncated at $m=200$ and $n=100$, to account for the size of the domain in different directions; beyond these, the solutions were indistinguishable as $m$ and $n$ increase. The central panel shows the flux through the mask at each point $Q(x,y)$, while the side panels show the flux out of the sides, given by the non-zero component of $\qvec$. The position of the mouth is shown by the red-dashed box. As the mouth is positioned at the centre of the mask, the flux out of the top and bottom, and the left and right of the mask are symmetric. We observe that  the flux through the mask varies spatially, with a larger flux closer to the mouth. This shows that treating a mask as a single resistor with uniform spatial flux, as in an equivalent circuit model, neglects important behaviour that is required for accurately predicting the flux. The figure also shows that the amount of leakage at each point on the boundary decreases as the distance from the mouth increases. This is because the gap width is uniform, and so the viscous resistance to flow in the gap is simply proportional to the distance the air flows through the gap. We also see that the leakage flux from the corners of the mask is zero. This arises from our assumption that the air everywhere outside the region covered by the mask is at atmospheric pressure and hence $\qvec$ becomes zero at the sharp corner at leading order (since $q_x=0$ along the horizontal boundaries and $q_y=0$ along the vertical boundaries in our approximation). We explore the effect of rounding at these corners in Appendix \ref{sec:Rounding}.

We note that, when $h=1$, so that $g(x,y)=f(x,y)+1$, it is possible to find $w$ explicitly, by substituting \eqref{eq:Asympt_SOL_pressure}, \eqref{eq:Asympt_Sol_velocity},   and \eqref{eq:Asympt_Reduced_gov_eqn} into \eqref{eq:Asympt_EQN_channel_mass}, then integrating with respect to $z$, and applying the boundary condition \eqref{eq:Asympt_BC_face}. We find that
\begin{equation}
    w(x,y,z) = (w_{\In}(x,y)-Q(x,y))\left( 2z^3 - 3(f+g)z^2 + 6fgz + f^3 - 3 f^2g \right) + w_{\In}(x,y).
\end{equation}

\subsection{Neck gaiters}
In the neck gaiter case, we solve \eqref{eq:Asympt_Reduced_gov_eqn} subject to the boundary conditions \eqref{eq:Asympt_Reduced_bc_periodic}, so that periodic conditions hold at $x=0$, $x=l$,
yielding similar expressions to \eqref{eq:2D_constant_series_Q}--\eqref{eq:2D_constant_series_q} as given by \eqref{eq:2D_constant_series_Q_gaiter}--\eqref{eq:2D_constant_series_q_gaiter} in Appendix \ref{sec:gaiter}. In Fig.~\ref{fig:2D_Example_Solution_Gaiter}, we  plot the flux of exhaled air from a neck gaiter, predicted by \eqref{eq:2D_constant_series_Q_gaiter}--\eqref{eq:2D_constant_series_q_gaiter}, when $ k =1$ in the case where the mouth is at the centre of the mask. As the material must wrap around the wearer's head, we assume the domain is larger and take $l=6$. From this solution, we  see that the majority of flux out through the neck gaiter material and the boundaries is occurring in the central portion of the domain, and so not all the material is being used. 

\begin{figure}
    \centering
    \includegraphics[width=0.97\textwidth]{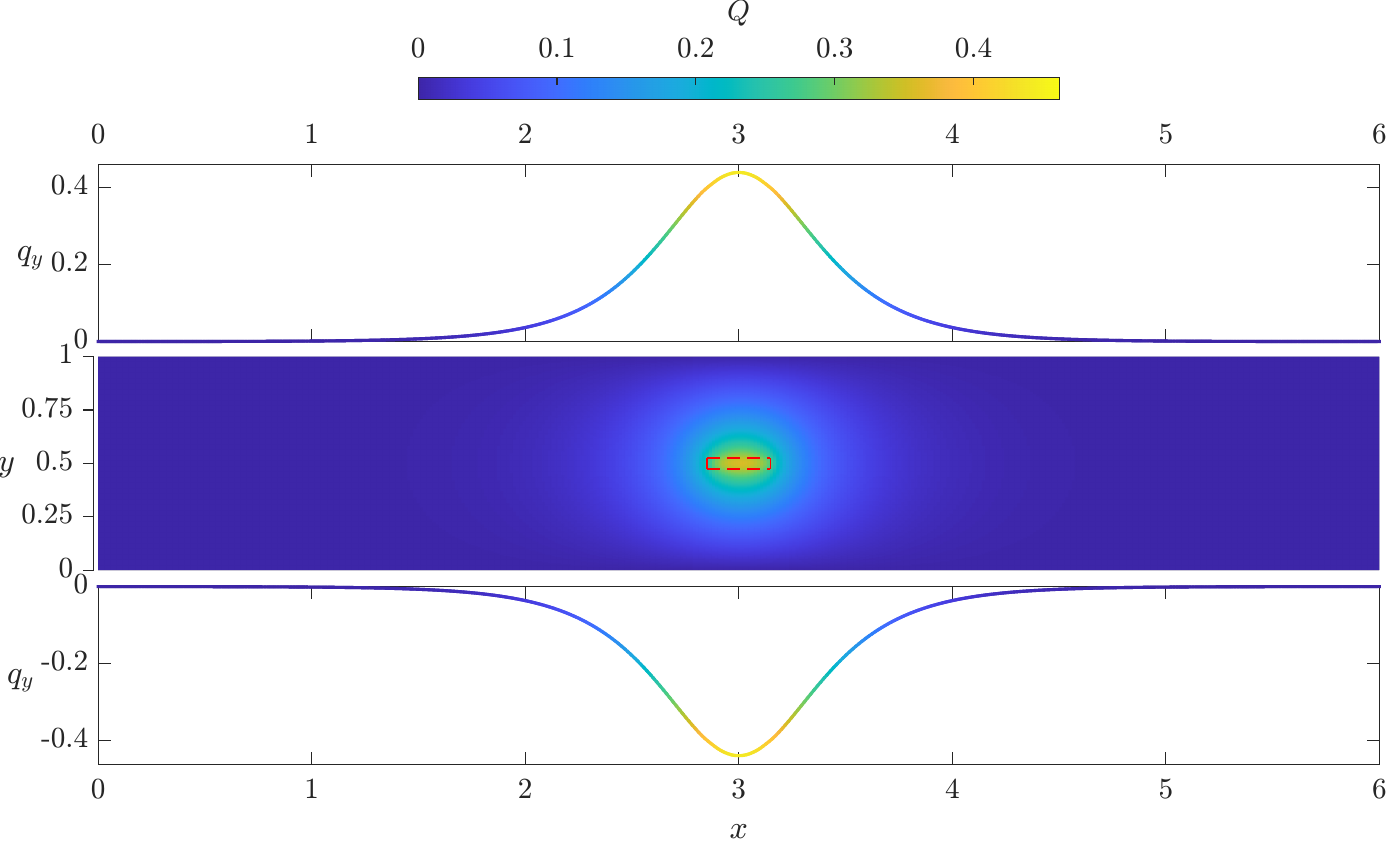}
    \caption{The central panel shows the flux $Q$ through a neck gaiter when $ k =1$. The airflow originates from  a mouth occupying $2.852 \le x \le 3.148$, $0.475 \le y \le 0.525$, shown as a dashed red line. The corresponding fluxes arising from leakage out of the side of the mask are shown in the surrounding panels. Colour is used to show the magnitude of the flux in all panels.}
    \label{fig:2D_Example_Solution_Gaiter}
\end{figure}

\begin{figure}[h]
    \centering
    \includegraphics[width=0.55\linewidth]{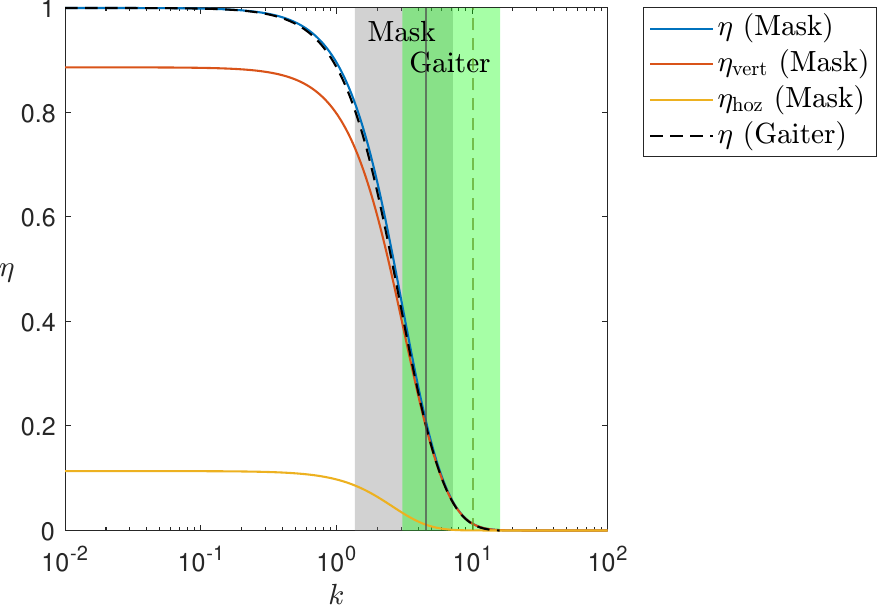}
    \caption{The leakage ratio $\eta$ for a face mask with predicted by the solution of \eqref{eq:Asympt_Reduced_gov_eqn}, \eqref{eq:Asympt_Reduced_bc} when $l=2$, shown as a blue line and the leakage ratio for a neck gaiter with $l=6$, predicted by the solution of \eqref{eq:Asympt_Reduced_gov_eqn}, \eqref{eq:Asympt_Reduced_bc_periodic}, shown as a black dashed line, both as  functions of $ k $. Both face coverings have constant gap widths and all other geometric parameters are kept identical. For the face mask we also show the horizontal and vertical leakage ratios, giving the proportion of the total flux out the horizontal and vertical boundaries, as red and yellow lines, respectively. The typical range of $ k $ for a face mask and a neck gaiter, based on the values in Table \ref{tab:dimensionless} are shown by a grey and a green region, respectively, and the values for a face covering with a typical fit are shown by a solid grey and a dashed green vertical line, respectively. }
    \label{fig:2D_Mask_NeckGaiter_Comparision}
\end{figure}

To compare the two mask designs, in Fig.~\ref{fig:2D_Mask_NeckGaiter_Comparision} we show the leakage ratio $\eta$ as a function of $ k $ for both a face mask with $l=2$ (as in Fig.~\ref{fig:2D_Example_Solution_Mask}) and for a neck gaiter with $l=6$ (as in Fig.~\ref{fig:2D_Example_Solution_Gaiter}), keeping all other geometric parameters identical. (We plot $\eta$ rather than $1-\eta$ for consistency with previous literature, \textit{e.g.} \cite{Ni2023}.)
We also show the horizontal leakage ratio $\eta_{\hoz}$ and the vertical leakage ratio $\eta_{\ver}$, defined in  \eqref{eq:Def_eta_hoz_vert} for the face mask. This allows us to see that the majority of the leakage occurs through the top and bottom boundaries. This is to be expected as these boundaries are closer to a mouth positioned in the centre of the mask, and as the gap width is constant, resistance to flow behind the mask is equal at every point.

Comparing the total leakage ratio $\eta$ for a face mask and a neck gaiter as a function of $ k $, we see that the two functions are almost indistinguishable. However, based on the parameter ranges for $K$ in Table \ref{tab:dimensionless}, we estimate the value of $ k $ for a face mask and a neck gaiter to be in \numrange{1.4}{7.2} and \numrange{3.1}{16.0}, respectively, depending on the fit. These ranges are shown in Fig.~\ref{fig:2D_Mask_NeckGaiter_Comparision} as grey and green regions, respectively, and the value of $k$ for a nominal fit are shown by grey and green vertical lines. We see that our model predicts that a face mask can leak as little as 5\% to as much as 81\% of the airflow, depending on the fit, with 21\% leakage with a typical fit, whereas the neck gaiter will have at most 40\% leakage, and only 1\% for a typical fit. This suggests that a well-fitted face mask may have less leakage than a badly fitted neck gaiter, but on average a neck gaiter will leak less than a face mask. However, our estimate of the gap width is based on simulations of the fit of a face mask, whose material is much less elastic than that of a face mask, so the worst-case leakage of 40\% arising from the gap width of an oversized mask is likely to be an overestimate. To see why $\eta$ is very close for a mask and neck gaiter when $ k $ is equal, we can compare Fig.~\ref{fig:2D_Example_Solution_Mask} and Fig.~\ref{fig:2D_Example_Solution_Gaiter}, which are both shown for $ k =1$. Although the neck gaiter removes the possibility of air escaping from the sides, this comes at the expense of larger upper and lower boundaries than the face mask from which air can leak out. As a result, approximately the same total leakage can occur in both the face mask and the neck gaiter.

\begin{figure}[h]
    \centering
    \includegraphics[width=0.6\linewidth]{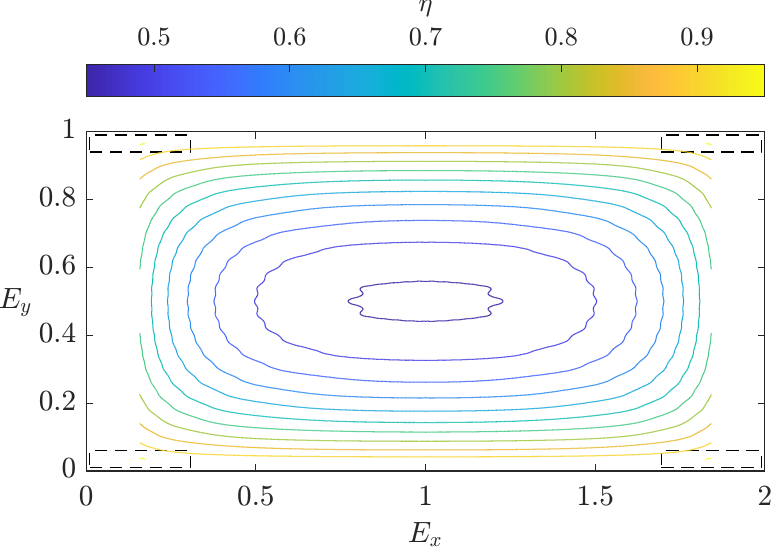}
    \caption{ The contours of the total leakage ratio $\eta$ for a face mask  predicted by the solution of \eqref{eq:Asympt_Reduced_gov_eqn}, \eqref{eq:Asympt_Reduced_bc} when $ k =3$ and $l=2$, as a function of the centre of a rectangular mouth $\left(E_{x},E_{y}\right)$, assuming the mouth has dimensions $\left( 2L_{x},2L_{y}\right) = \left(0.296,0.05\right)$.  The contours show increases in graduations of 0.05 from 0.45 to 0.95. Example mouth shapes at the corners are shown  by black dashed lines.}
    \label{fig:2D_MouthPosition}
\end{figure}

Until now, we have only considered the amount of leakage when the mouth is positioned in the centre of the mask.      
In Fig.~\ref{fig:2D_MouthPosition}, we show how the total leakage ratio varies as the centre of the mouth, $\left(E_{x},E_{y}\right)$, varies across the $xy$-plane. The results show that the lowest leakage occurs when the mouth is positioned at the centre of the mask. This is expected as it maximises the distance from the mouth to the sides of the mask. We also see that the increase in leakage ratio as the mouth moves towards the edge is nonlinear, with small perturbations from the centre having a small effect on the leakage, but a large increase in the leakage when the mouth approaches the edge of the mask. We could also consider varying the shape of the mouth, but we find this has minimal effect on the amount of leakage (see Appendix \ref{sec:App_shape} for details).

\subsection{Two-dimensional results: variable gap width}

\begin{figure}[htbp]
    \centering
    \begin{minipage}[b]{0.25\linewidth}
        \begin{overpic}[trim={4cm 0 5.3cm 0},clip,width=\linewidth]{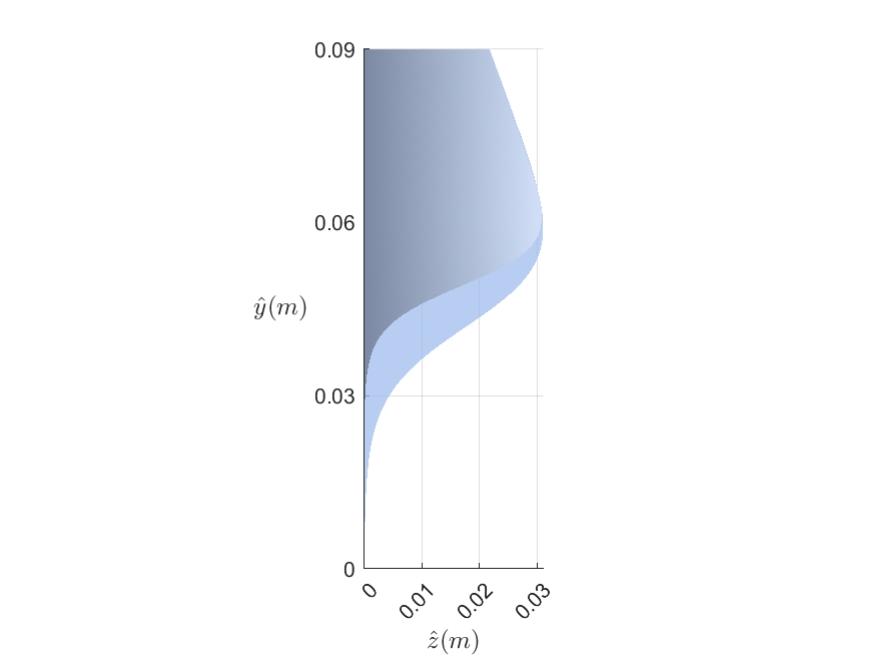}
        \put(3,94){\textbf{(a)}}
        \end{overpic}
        \vspace{1cm}
    \end{minipage}
    ~~~
    \begin{minipage}[b]{0.6\linewidth}
        \begin{overpic}[trim={0 3cm 1cm 4cm},clip,width=\linewidth]{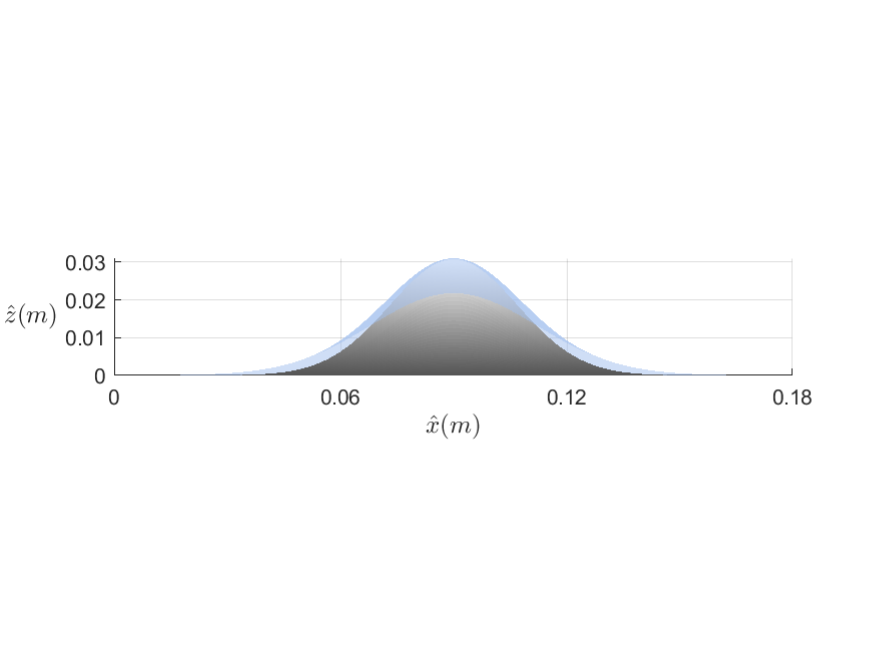}
        \put(0,30){\textbf{(b)}}
        \end{overpic}
        \begin{overpic}[trim={0 0.3cm 0.3cm 1.2cm},clip,width=\linewidth]{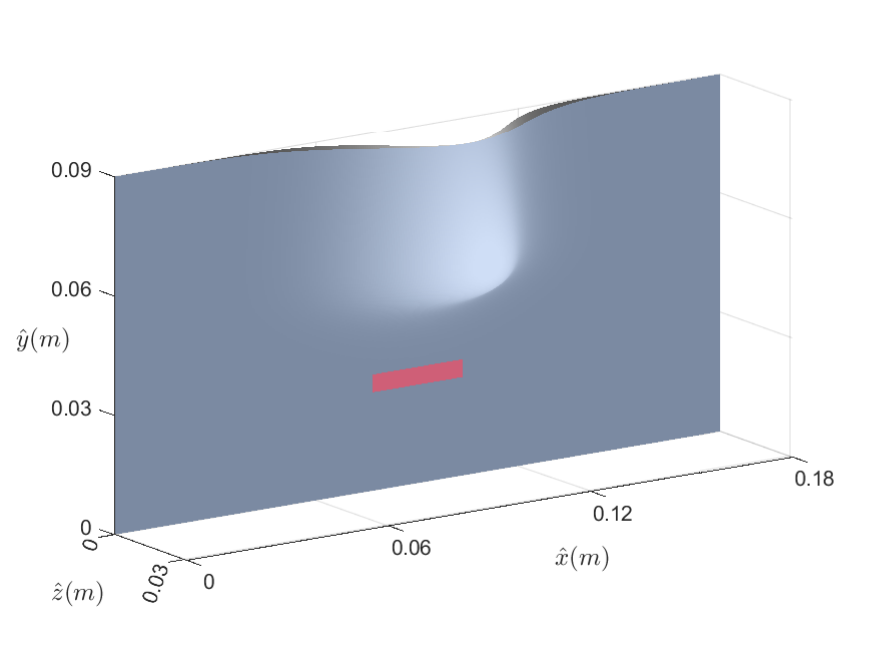}
        \put(0,65){\textbf{(c)}}
        \end{overpic}
    \end{minipage}
    \caption{Graphics showing the dimensional shape of the face (grey surface), and the position  of the back of the mask, shown as a transparent blue surface. We show in (a) the side view, (b) the top view, and (c) the three dimensional view.  The location of the mouth is shown by a red rectangle.}
    \label{fig:NoseFunction}
\end{figure}

\begin{figure}[htbp]
    \centering
    \begin{overpic}[width=0.6\linewidth]{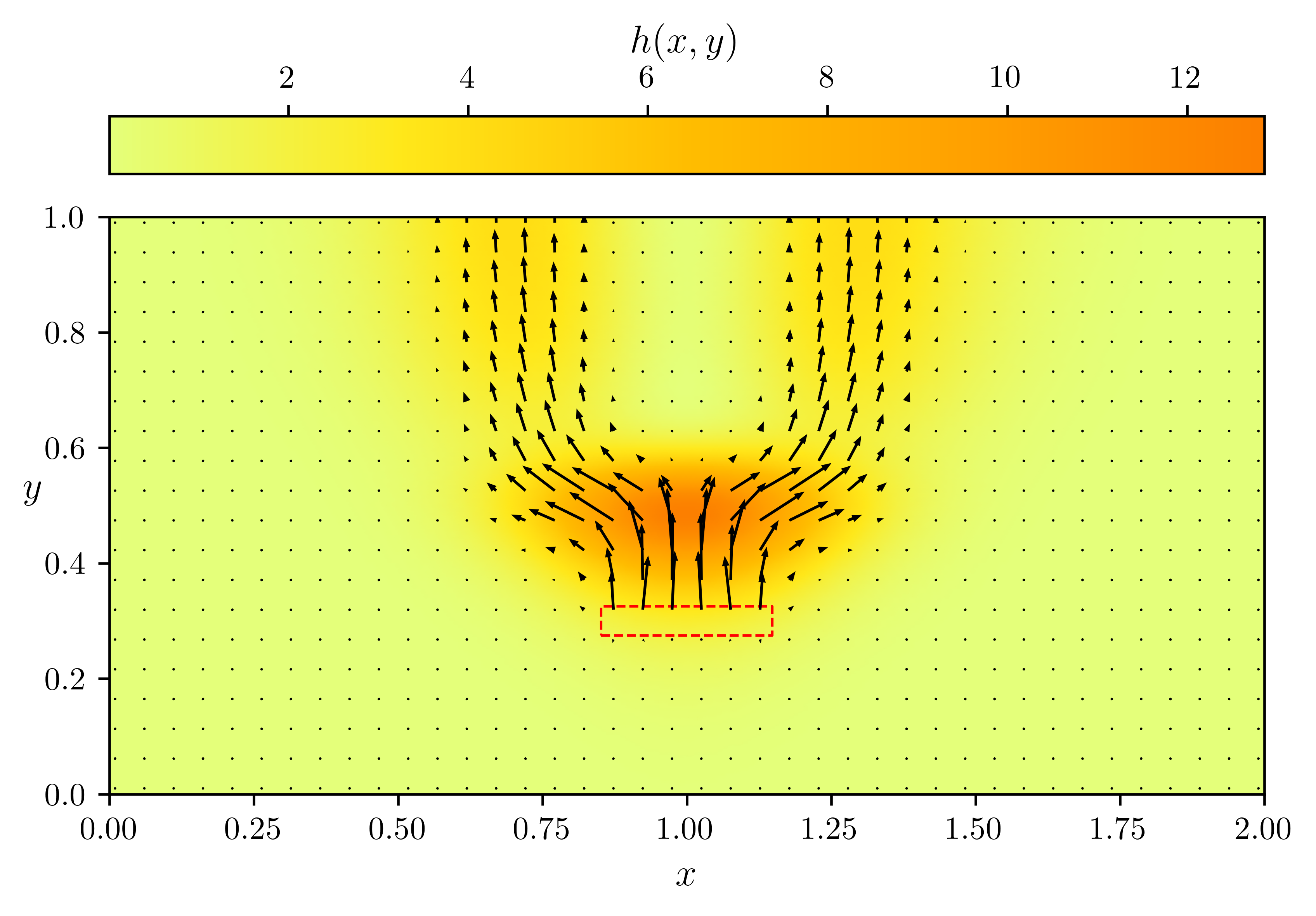}
        \put(-20,70){\textbf{(a)}}
    \end{overpic}
    \hfill
    \begin{overpic}[width=0.9\linewidth]{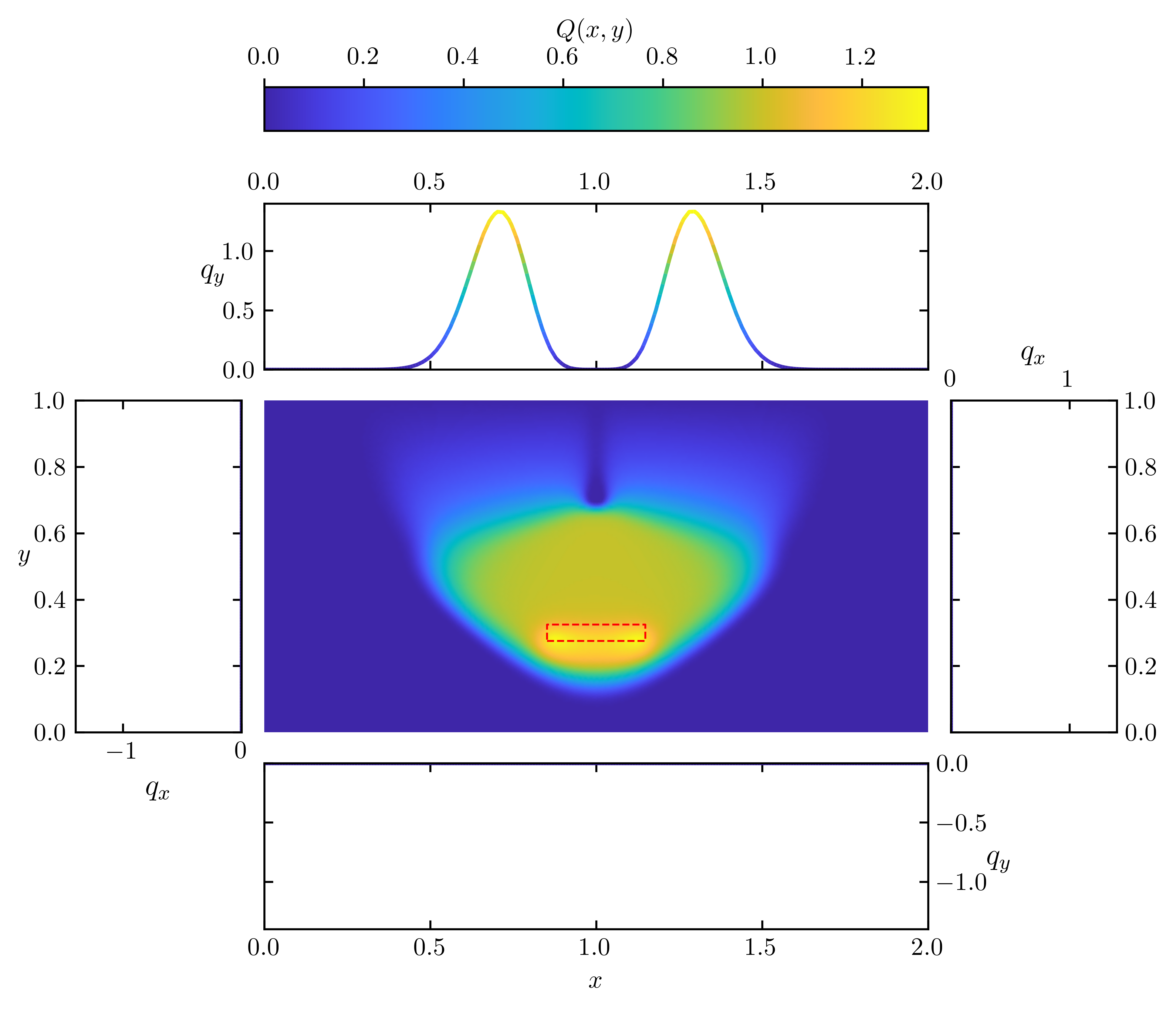} 
        \put(3,85){\textbf{(b)}}
    \end{overpic}
    \caption{(a) The gap width $h(x,y)=g(x,y)-f(x,y)$ with $f$ and $g$ given by \eqref{eq:Variable_F} with the parameter values given in Table \ref{tab:geometry_parameters} is shown as a heat map. The depth-integrated flux $\qvec$  obtained by solving the reduced model \eqref{eq:Asympt_Reduced_gov_eqn}--\eqref{eq:Asympt_Reduced_bc} when $k=6.0$ is shown by black arrows, with the boundary of the region occupied by the mouth $0.852 \le x \le 1.148$, $0.275\leq y \leq 0.325$, shown by a red-dashed line. (b) The flux through the mask $Q(x,y)$ from the same solution is shown in the central panel with the flux leaking out of the sides shown in the side panels. }
    \label{fig:2D_Example_Solution_variable_h}
\end{figure}

\begin{figure}
    \hfill
    \includegraphics[width=0.8\linewidth]{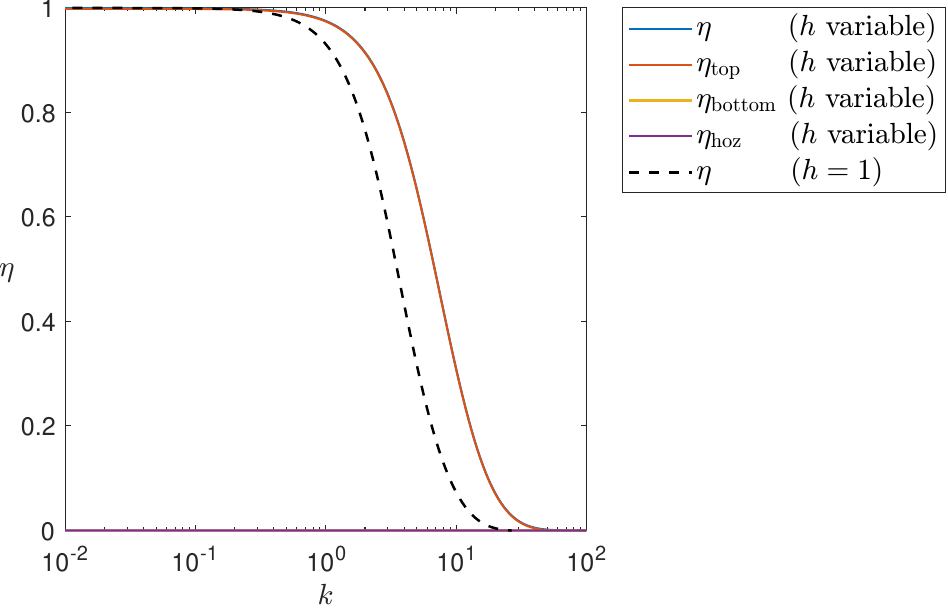}
    \caption{The leakage ratio $\eta$ of a face mask with a variable gap width $h(x,y)$ with unit mean, predicted by the solution of the reduced model \eqref{eq:Asympt_Reduced_gov_eqn}--\eqref{eq:Asympt_Reduced_bc} as a function of $k$ is shown as a solid blue line. The top, bottom, and horizontal leakage ratios are shown as red, yellow and purple lines, respectively. The leakage ratio predicted by the solution of \eqref{eq:Asympt_Reduced_gov_eqn}--\eqref{eq:Asympt_Reduced_bc} with constant $h\equiv1$ is shown by a black dashed line.}
    \label{fig:2D_Leakage_Ratio_Sweep_Variable_height}
\end{figure}

In reality, the width of the gap between a face covering and the wearer's face is not constant, so to predict realistic behaviour we must solve \eqref{eq:Asympt_Reduced_gov_eqn}--\eqref{eq:Asympt_Reduced_bc} for variable $h$. We do this by employing the finite element method  (also used  in Section \ref{sec:Leakage_2D_constant} to solve the reduced model with constant $h$ for a mask with rounded corners). We will solve for a given functional form for $h$. We note that an analytic expression is not necessary and the problem could be solved for a given set of measurements \cite{Hariharan2021} or using the output from of simulations \cite{Buxton2020,Cai2016,Lei2014,Lei2012headform,Solano2021}.    We focus on using simple functions to obtain qualitatively correct results. We define   $f(x,y)=F\left(x,y,\lambdavec^{f}\right)$, and $g(x,y)=F\left(x,y,\lambdavec^{g}\right)$ where $\lambdavec^{f}$ and $\lambdavec^{g}$ are different sets of parameters and the function $F\left(x,y,\lambdavec\right)$ for $\lambdavec=\lambdavec^f$ or $\lambdavec^g$ is defined by 

\begin{equation}
    F(x,y; \lambdavec) = \exp\left( - \frac{\left( x- \lambda_{1}\right)^2}{\lambda_{2}+\lambda_{3}y} \right) \frac{\lambda_{4} \left(1-\lambda_{5}y\right)}{1 + \exp\left( - \lambda_{6}\left( y - \lambda_{7}\right)\right)}+\lambda_{8},\label{eq:Variable_F}
\end{equation}
where $\lambdavec=\left(\lambda_{i}\right)_{i=1}^{8}$. The function $F$ consists of a Gaussian distribution in the $x$-direction multiplied by a function representing the silhouette of a nose in side profile. The parameter $\lambda_{1}$ controls the horizontal position of the nose, and $\lambda_{2}$ and $\lambda_{3}$ control the horizontal spread, with $\lambda_{3}$ controlling the variation of the spread in the $y$ direction. The parameter  $\lambda_{4}$ controls the maximum size of the nose, and $\lambda_{5}$ controls the rate of decrease from the tip to the bridge of the nose. The parameters $\lambda_{6}$ and $\lambda_{7}$ control the position of the bottom of the nose and the rate at which it grows to the tip, and $\lambda_{8}$ allows for an offset at the edge of the mask. The values of $\lambdavec^{f}$ and $\lambdavec^{g}$ used throughout this paper are given in Table~\ref{tab:geometry_parameters} and chosen such that $h(x,y)=g(x,y)-f(x,y)>0$ for all $(x,y)\in\Omega$, {and} the mean of $h$ is one.  Using the scalings \eqref{eq:Nondim_Scalings}, we find that the choice of parameters for $f$ gives a nose extending \SI{3}{cm} from the face, while the choice of parameters for $g$ gives a gap size either side of the nose peaking at \SI{4}{mm}, which corresponds to the maximum gap size around the nose in the profiles given in \cite{Ni2023}. The resulting dimensional profile is shown in Fig.~\ref{fig:NoseFunction}.

\begin{table}[h!]
\caption{Parameter values so that \eqref{eq:Variable_F} approximates the shape of face $f$ and a face mask $g$.  } 
\label{tab:geometry_parameters}
\begin{ruledtabular}
\begin{tabular}{ccccccccc}
         & $\lambda_{1}$ & $\lambda_{2}$& $\lambda_{3}$& $\lambda_{4}$& $\lambda_{5}$& $\lambda_{6}$ & $\lambda_{7}$ & $\lambda_{8}$\\ \hline
        $\lambdavec^{f}$ & $1.0$ & $0.07$& $0.0$ & $50.0$& $0.6$& $25.0$& $0.55$ & $0.0$ \\
        $\lambdavec^{g}$ & $1.0$ & $0.02$& $0.1$ & $51.0$& $0.605$& $15.0$& $0.48$ & $0.01$
\end{tabular}
\end{ruledtabular}
\end{table}

In Fig.~\ref{fig:2D_Example_Solution_variable_h}a, we show a heat map of $h(x,y)$ using the parameter values given in Table~\ref{tab:geometry_parameters}. We see that there is a small `u'-shaped region of gap width $h\approx {2.0}$ due to the nose (which joins the mouth to the top boundary) surrounding a region of very small gap width representing close fitting over the nose. We also show in Fig.~\ref{fig:2D_Example_Solution_variable_h}a 
the lateral flux in the gap behind the mask, $\qvec$, (as black arrows), in the case when $k=6.0$, obtained by solving \eqref{eq:Asympt_Reduced_gov_eqn}--\eqref{eq:Asympt_Reduced_bc} 
and using \eqref{eq:Asympt_sol_q}, along with the position of the mouth as a red dashed line. We keep our attention focused on the flow from the mouth rather than adding in complications from the additional nasal flow.
We see that the majority of the airflow is upwards in the gap along the `u'-shaped channel, with the speed decreasing away from the mouth, and as $h$ decreases. In  Fig.~\ref{fig:2D_Example_Solution_variable_h}b, we show the corresponding flux through the mask $Q$, and the leakage out of the sides. The position of the mouth is again shown as a red dashed box. We see that the flow through the `u'- shape channel creates a double-peaked distribution in the flow out of the top of the mask, and the presence of the nose creates an asymmetry in the flow {through the mask}, as expected. The flux through the mask is approximately uniform above the  large-gap-width region located below the nose. This is because the gap is so large there is little resistance to the flow at that location, so it is easier for the air to spread out. 
We also see that the peak (\textit{i.e.} the {bright} yellow region) of the flow rate through the mask is located beneath the mouth, rather than centred on it as seen in the constant-width solution shown in Fig.\ref{fig:2D_Example_Solution_Mask}. Physically, we can interpret this as air travelling along the side of the nose finding less resistance to leaking out of the sides compared with flowing through the mask, whereas the flow over the chin finds less resistance when flowing through the mask compared with leaking out.

In Fig.~\ref{fig:2D_Leakage_Ratio_Sweep_Variable_height}, we  present the leakage ratio for the variable gap width case as $k$ varies. As our choice of mouth position and gap width is symmetric horizontally, we also show $\eta_{\textrm{top}}$, $\eta_{\textrm{bottom}}$, and $\eta_{\textrm{hoz}}$. The results show the same qualitative behaviour as the constant-$h$ case, with $\eta \to 1$ as $k\to0$, and $\eta\to0$ as $k\to\infty$, with a transition as $k$ goes between approximately $0.1$~and~$10$. The different components of the leakage ratio show, as expected from Fig.~\ref{fig:2D_Example_Solution_variable_h}, that all of the leakage occurs out of the top boundary for all $k$. In the figure, we also show the leakage ratio predicted by the constant-gap-width model, when $h\equiv1$, the mean of the variable-gap-width model. We see that the constant-width model gives a reasonable approximation for the {shape of the} variable-width curve, but at lower values of $k$. This discrepancy gives a difference in predicted values up to 30\%.  
These results suggest that, although the constant-width model cannot accurately predict where leakage will occur around the perimeter of a mask, it gives a reasonable approximation of the more complicated model provided the gap width is correctly parametrised. The sweep presented in Fig.~\ref{fig:2D_Leakage_Ratio_Sweep_Variable_height} was computed for 250 data points and took only \SI{50}{s} to run on a standard desktop, so each simulation takes only \SI{0.2}{s} on average. Individual simulations take \SI{2.5}{s} on average, suggesting mesh creating is much more costly than the problem solution. Therefore, even with variation in $h$, our reduced model is six orders of magnitude faster to run than any of the CFD simulations described in the Introduction.

\section{Aerosol transport}\label{sec:Efficiency}

We now solve for the leading-order steady aerosol concentration. The leading-order  transport equation in the gap \eqref{eq:NondimEQNS_gap_conc} is 
\begin{equation}
    \uvec \bcdot \bnabla c = 0,\label{eqn:cb1}
\end{equation}
which we solve subject to the boundary condition
\begin{equation}
    c = 
    \begin{cases}
        1 & \text { if } (x,y) \in \OmegaMouth\\
        0 & \text { if } (x,y) \notin \OmegaMouth
    \end{cases}
\end{equation}
on $z=f(x,y)$. Equation \eqref{eqn:cb1} implies that $c$ is constant along streamlines and, since all streamlines originate from the mouth, it follows that $c=1$ everywhere in the gap and, in particular, on $z=g(x,y)$. The leading-order form of the governing equation of concentration inside the face covering \eqref{eq:NondimEQNS_mask_conc} is 
\begin{equation}
    \frac{\partial C}{\partial z} = - \Gamma C,
\end{equation}
subject to $C=1$ at $z=g(x,y)$ from \eqref{eq:NondimBC_mask_rear}. 
The solution is
\begin{equation}
    C = e^{-\Gamma(z-g)}.
\end{equation}
We straightforwardly calculate the filtration efficiency, to find that
\begin{equation}
    \FE = 1 - \dfrac{\displaystyle\int_{0}^{1}\int_{0}^{l}Q(x,y)C(g(x,y)+T)\,\mathrm{d}x\,\mathrm{d}y}{\displaystyle\int_{0}^{1}\int_{0}^{l}Q(x,y)C(g(x,y))\,\mathrm{d}x\,\mathrm{d}y} = 1- e^{-\Gamma T},\label{eq:Asympt_FE}
\end{equation}
which has equivalent form to the formula given in \cite{Zangmeister2020filtration}. Since we are considering an advection-only model, it is straightforward to see that the face-fitted filtration efficiency
\begin{equation}
    \FFE = \left( 1- \eta \right)\FE.\label{eq:Asympt_FFE}
\end{equation}
    
So far, we have assumed that the permeability and capture rate of a material are independent properties. However, we would intuitively expect a less permeable material to have a higher capture rate, as it will contain more microscale obstacles that impede the aerosol particles. To obtain an expression that captures this relationship, we assume that the permeability is related to the porosity, $\phi$, through the Kozeny--Carmen equation \cite{NieldBejan}, a commonly used empirical model, in which
\begin{equation}
    \hat{K} = \frac{\hat{D}_{p}^2\phi^3}{180\left(1-\phi\right)^2},
\end{equation}
where $\hat{D}_{p}$ is the effective average diameter of the microscale fibers (\unit{m}).
Assuming the capture rate $\hat{\gamma}$ to be proportional to the available surface area on the microstructure, it is straightforward to deduce that 
\begin{equation}
    \hat{\gamma} = \hat{\gamma}_{0}\left( 1- \phi\right)^{2/3},\label{eq:Capture_EmpiricalLaw}
\end{equation}
where $\hat{\gamma}_{0}$ (\unit{m^{-1}}) is the surface reaction rate constant, which depends on 
the size and shape of the fibres and their packing and any coatings applied \cite{VirustaticMaterial2021}. 
The $\phi$-dependent versions of $k$ and $\Gamma$ are
\begin{equation}
    k = k_{0} \frac{\phi^{3/2}}{1-\phi}, \quad \Gamma = \Gamma_{0} \frac{\left(1-\phi \right)^{2/3}}{\phi},
\end{equation}
$k_{0}=\hat{D}_{p}/\left(\sqrt{15}\epsilon^2 \LgthScle \right)$ is a shape parameter that captures the effect of the microscale geometry and $\Gamma_{0}=\hat{\gamma}_{0}\LgthScle$ is the dimensionless surface reaction rate constant, capturing the effect of the surface chemistry on the macroscale capture rate. Combining these expressions allows us to link the capture number to the dimensionless permeability, which, for convenience, we write in terms of~$k$ as
\begin{equation}
    \Gamma = \frac{\Gamma_{0}k_{0}^{2/3}}{k^{2/3}}.\label{eq:Empirical_Gamma_k}
\end{equation}
Relationship \eqref{eq:Empirical_Gamma_k} has the property that, as the permeability becomes infinite, the capture number tends to zero, whereas if the permeability tends to zero then the capture rate becomes infinite. The derivation of these expressions could be made more formal by specifying a particular microstructure model and geometry, then applying the method of multiple scales \cite{Dalwadi2015}, but our expressions exhibit the correct qualitative behaviour, so we do not pursue this further here. 

\begin{figure}
    \centering
    \includegraphics[width=\linewidth]{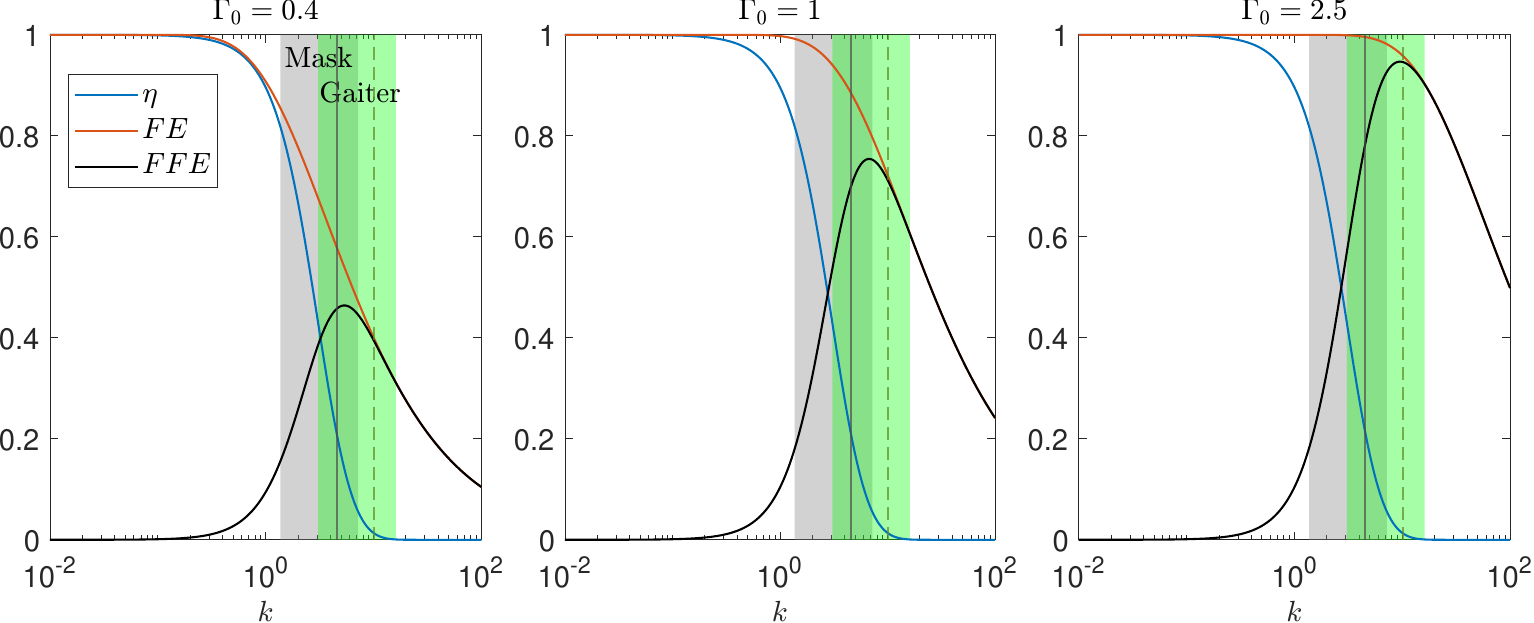}
    \caption{The leakage ratio $\eta$ (blue), the filtration efficiency $\FE$ (red) and face-fitted filtration efficiency $\FFE$ (black) as functions of $k$ when $k_{0}=14.52$ and (a) $\Gamma_{0}=0.4$, (b) $\Gamma_{0}=1.0$,(c) $\Gamma_{0}=2.5$, predicted by the solution of \eqref{eq:Asympt_Reduced_gov_eqn}, \eqref{eq:Asympt_Reduced_bc} and \eqref{eq:Asympt_FE} when $h=1$, $l=2$, and the mouth occupies $0.852 \le x \le 1.148$, $0.475 \le y \le 0.525$. The typical range of $ k $ for a face mask and a neck gaiter, based on the values in Table \ref{tab:dimensionless} are shown by a grey and a green region, respectively, and the values for a face covering with a typical fit are shown by a solid grey and a dashed green vertical line, respectively.  }
    \label{fig:Optial_FFE_k}
\end{figure}

\begin{figure}
    \centering
    \includegraphics[width=0.7\textwidth]{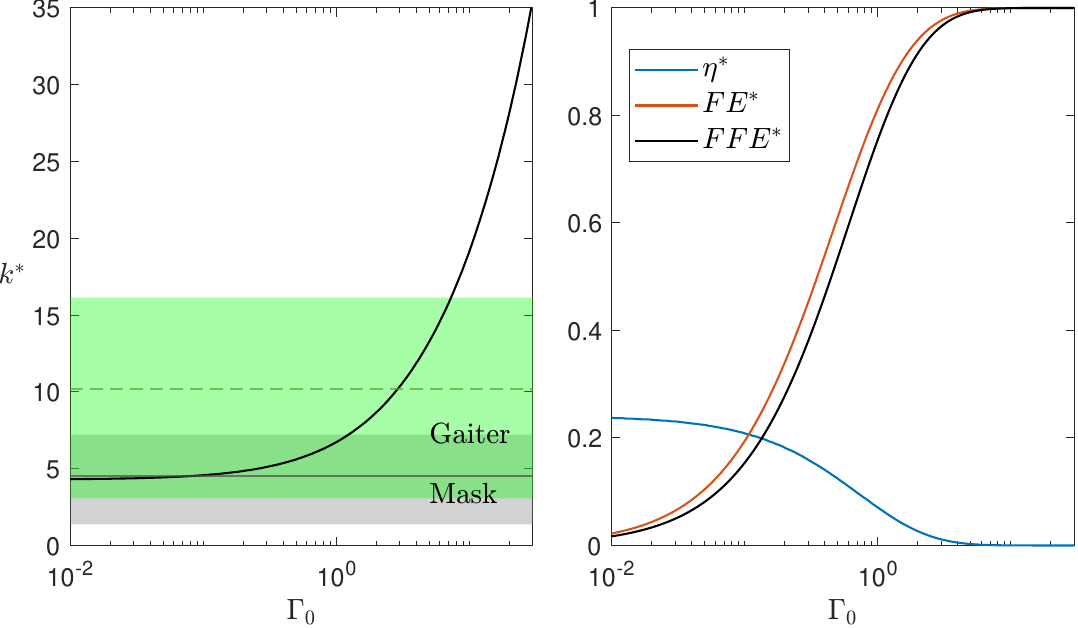}
    \caption{(Left) The value of $ k $ that maximises $\FFE$ as a function of the dimensionless surface reaction rate constant $\Gamma_{0}$. (Right) The maximum value of $\FFE$ (black), and the corresponding values of $\eta$ (blue) and $\FE$ (red). In both panels the values are calculated using the solution of \eqref{eq:Asympt_Reduced_gov_eqn}, \eqref{eq:Asympt_Reduced_bc} and \eqref{eq:Asympt_FE}--\eqref{eq:Asympt_FFE} when
    $k_{0}=14.52$, $l=2$, and the mouth occupies $0.852 \le x \le 1.148$, $0.475 \le y \le 0.525$. The typical range of $ k $ for a face mask and a neck gaiter, based on the values in Table \ref{tab:dimensionless} are shown by a grey and a green region, respectively, and the values for a face covering with a typical fit are shown by a solid grey and a dashed green vertical line, respectively. }
    \label{fig:Optimal_FFE_capture_rate}
\end{figure}

In Fig.~\ref{fig:Optial_FFE_k}, we use \eqref{eq:Empirical_Gamma_k} to show how $\eta$, $\FE$ and $\FFE$ vary with $k$, with each panel showing a different value of $\Gamma_{0}$. For simplicity, we use the values of $\eta$ obtained by solving the reduced model for a face mask, \eqref{eq:Asympt_Reduced_gov_eqn}--\eqref{eq:Asympt_Reduced_bc}, assuming constant gap width $h\equiv 1$, and a rectangular mouth at the centre of the mask. We see that $\eta$ is the same in each of the panels, since $\eta$ is independent of $\Gamma_0$. We do not show an equivalent curve for a neck gaiter, since, as already discussed, $\eta$ as a function of $ k $ is indistinguishable for a face mask and a neck gaiter. The ranges of $ k $ for a face mask and a neck gaiter, as the fit of the face covering changes, are shown as grey and green regions, respectively, with the value of a typical fit shown by vertical lines. All the panels show the existence of an optimal value, $k^*$ that gives the maximum $\FFE^*$. It is clear that, as $\Gamma_{0}$ increases, the values of $k^{*}$ and $\FFE^{*}$ both increase, so that the optimal point provides greater protection and is more permeable and/or tighter fitting as the surface reaction rate constant increases. We also see that, at lower values of $\Gamma_{0}$, $k^{*}$ lies in the range of possible values for both a face mask and a neck gaiter (with different mean gap widths). However, as  $\Gamma_{0}$ increases, the optimum moves out of the range of $k$s possible for a face mask. Comparing the typical values of $k$ shown by the vertical lines in Fig.~\ref{fig:Optial_FFE_k}, we see that, for low $\Gamma_{0}$, the face mask is less permeable than the optimum and the neck gaiter is more permeable than the optimum, and thus both under-filter the aerosols compared with the optimum. However, as $\Gamma_{0}$ increases, the value of $k$ that gives the optimal efficiency moves away from the mean value associated with face masks and gets much closer to the mean value for neck gaiters. We illustrate how $k^{*}$ and $\FFE^{*}$, along with the corresponding leakage ratio $\eta^{*}$ and filtration efficiency $\FE^{*}$, vary with $\Gamma_0$ in
in Fig.~\ref{fig:Optimal_FFE_capture_rate}. We see that, to achieve a face-fitted filtration efficiency above 90\%, the leakage ratio must be less than 2\%.

The range of possible values of $ k $ for a face mask and a neck gaiter are also shown in Fig.~\ref{fig:Optimal_FFE_capture_rate}, and suggest that, by adjusting the fit, the optimal value $k^{*}$ can be achieved by a face mask for $\Gamma_{0}<1.2$, whereas a neck gaiter can achieve the optimal value for $\Gamma_{0}<7.2$. This means that, with these parameter values, our model predicts that a face mask can have a face-fitted filtration efficiency of at most 80\%, whereas a neck gaiter can have a face-fitted filtration efficiency up to 99\%. It is reasonable to also ask about the dependence of $\FFE$ on the only other remaining parameter in the problem, $k_{0}$. However, the results presented depend on $k_0$ only through \eqref{eq:Empirical_Gamma_k} and hence variation in $k_{0}$ is equivalent to variation in $\Gamma_{0}$.

\section{Conclusions, discussion, and extensions}\label{sec:Discussion}

In this paper, we have built and solved a model for the airflow and aerosol transport in and behind a face mask and a neck gaiter during exhalation. By working in the asymptotic limit of small aspect ratio of air gap to mask size, and small reduced Reynolds number, we were able to obtain a reduced model \eqref{eq:Asympt_Reduced_gov_eqn}--\eqref{eq:Asympt_Reduced_bc_periodic} for the airflow. This model was considerably easier to solve than a full CFD simulation, whilst retaining the effect of spatial variation that is not captured in previous reduced models. We used this model to understand how the various parameters in the problem affect two key real-world performance metrics: the leakage ratio $\eta$, and the face-fitted filtration efficiency $\FFE$, previously identified in the literature \cite{Mittal2023Review}. The behaviour of this reduced model was determined principally by a key dimensionless parameter, $k$, given by $k^2 = 12 \LgthScle^2 \hat{K}/\left( \hat{h}_{0}^3 \hat{T}\right)$,  where $\hat{K}$ is the permeability of the material of the face covering, and $\hat{h}_{0}$ is a reference value for the size of the gap between the face covering and the wearer's face. Thus, $k^2$ is the ratio between the resistance to flow through the mask and the resistance to flow through the gap between the face and the face covering.

For the simplest case of a constant gap between face covering and the wearer's face, it was possible to solve \eqref{eq:Asympt_Reduced_gov_eqn}--\eqref{eq:Asympt_Reduced_bc_periodic} explicitly, and find expressions for the leakage ratio for both a mask and a neck gaiter. Our results show that, as $ k  \to 0$ (i.e. $\hat{K}\to0$, or $\hat{h}\to \infty$), $\eta \to 1$, whereas as $ k  \to \infty$ (i.e. $\hat{K}\to\infty$, or $\hat{h}\to 0$), $\eta \to 0$. In practice, this transition from all the flow leaking out of the sides to all the flow passing through the mask occurs as $ k  $ ranges from approximately $0.1$ to $10$. Besides $ k $, the other parameters in the reduced model relate to the geometry of the mouth. We found that the mouth shape and position only had a small effect on the amount of leakage for any plausible choice of geometry. Therefore, the leakage ratio is determined almost entirely by the value of $ k $. Using appropriate values for the permeability, we were able to estimate the possible range of $ k $ for both a surgical face mask and a neck gaiter, which suggests that a neck gaiter will have less leakage for the same gap width. However, as neck gaiters are made of a more elastic material, they may conform better to the face and create a smaller gap. Both of these factors will increase $ k $, and hence lead to a lower leakage ratio, highlighting the potential superiority of neck gaiters over face masks in minimising viral spread.

We also solved \eqref{eq:Asympt_Reduced_gov_eqn}--\eqref{eq:Asympt_Reduced_bc_periodic} numerically (using finite element methods) in the more general case of a variable gap size. Each solution is computed in less than \SI{2.5}{s}, representing a very significant saving in time and computational resources compared with a CFD simulation. By choosing an appropriate functional form for the gap size, we were able to reproduce realistic leakage profiles with a channel of larger gap size connecting the mouth to either side of the nose. We saw that the leakage rates depended on the dimensionless permeability in a qualitatively similar manner to that for a constant gap width, and was reasonably well approximated by the constant gap width model using the mean of the variable gap width. This represents a step improvement over previous reduced models in the literature, which could not capture the variation in the gap size away from the perimeter of the mask.

Our results show that increasing the permeability of the material will increase the proportion of air that flows through the face covering. However, increasing the permeability is also likely to decrease the filtration efficiency of the material, as there will be less material to impede the particles. To investigate this, we developed a simple model for the advection of aerosol particles by the airflow, with a capture rate that depends linearly on the local concentration of aerosol particles, the cross-sectional area of the material microstructure, and a surface reaction rate constant. By using the empirical Kozeny-Carman equation, we related the capture rate to the permeability of the material. Using this relationship, for a face covering of constant gap width we predicted the existence of an optimal material permeability which maximises the face-fitted filtration efficiency. By varying the surface rate constant of the material, we were able to show that the optimal value of face-fitted filtration efficiency always requires a face covering with only a small amount of leakage. These results also show that increasing the surface rate constant, for example by applying a coating \cite{VirustaticMaterial2021}, can substantially increase the maximum value of $\FFE$ and increase the corresponding value of $ k $ at which the maximum occurs,  to the point where they can be achieved by only a neck gaiter not a face mask.

Throughout this paper, we have considered the performance of the face covering during exhalation. However, by applying the  transformation $Q\to -Q$ to \eqref{eq:Asympt_Reduced_gov_eqn}--\eqref{eq:Asympt_Reduced_bc}, it is easy to see that leakage ratios predicted by our models for constant exhalation also describe the proportion of the flux through the mask during constant inhalation (since we are in a Stokes' flow regime). Further, by assuming that there is a constant background concentration of aerosol particles outside the mask, it is again straightforward to show that the appropriately re-defined $\FE$ and $\FFE$ for inward protection \cite{Mittal2023Review} correspond exactly to the expressions we predict for exhalation. However, experimental results in \cite{Si2024visualization} may be used to suggest that a small difference exists between $\eta$ for inhalation and exhalation;  either the effect of the flow on the gap size, or inertial effects must distinguish one process from another.

In future, it would be interesting to consider a more detailed particle transport model accounting for additional physical processes, such as inertia, molecular diffusion, and electrostatic forces, as these can all be important for different-sized aerosol particles \cite{Mittal2023Review}. It would also be interesting to employ systematic upscaling techniques, such as in \cite{Dalwadi2015}, to derive effective macroscale equations for the behaviour of the mask, where the macroscale parameters depend on microscale geometry and properties, such as weave pattern, thread size, and chemistry. The model could also be formulated in curvilinear coordinates to better capture the underpinning curvature of the head. However, we anticipate that, as the size of the gap is small compared with the curvature of the head, we would end up with the same model \eqref{eq:Asympt_Reduced_gov_eqn}--\eqref{eq:Asympt_Reduced_bc_periodic}, since the same result is found for the standard lubrication flow equations \cite{OckendonPDEs}. 
In our analysis, we focused on exhalation from the mouth and neglected any nasal flow. However, none of our calculations require the region where air exits the face to be a connected region, and so the model is also valid for exhalation from disjoint nostrils, albeit with a different $\OmegaMouth$. In particular, for a face mask with constant gap width $h$, the expression for $\beta_{mn}$ \eqref{eq:2D_constant_series_beta_general} may be split into integrals over each discrete region, allowing, for example, the solution for two rectangular nostrils to be calculated using \eqref{eq:2D_constant_series_beta_rectangle} for each nostril. An identical result also holds for neck gaiters. In our model, assume that the face covering material is inextensible. The model may be naturally extended to include an elastic face covering with permeability that may vary with tension. Spatial variations in permeability might affect the leakage calculation in interesting ways. We note that, inevitably, the constant-thickness case will be less relevant when we consider extensible materials. 
Another key extension will be to consider dynamic breathing (i.e. $w_{\In}$ is a function of time) and mouth opening and closing ($\OmegaMouth$ is a function of time). This has the potential to alter $h$ dynamically, especially if combined with the elastic model. Finally, this model was derived for velocities arising for light breathing, so it would be interesting to compare it with numerical simulations incorporating inertia in the airflow, which will become important during talking and coughing.

Our model provides a systematic framework for scientists and innovators to study permeable face masks and to determine how their operation depends on their physical parameters, aerosol capture rate, and fit. This will enable the expensive and time-consuming experiments to be focused on the relevant parameter regimes. It is a stepping stone towards re-evaluating how face masks protect the population from viruses released by infected wearers, which goes beyond current legislation, providing the opportunity for minimising transmission in future pandemics. 

\section{Acknowledgements}
The main part of this work was funded by an Innovate UK Analysis for Innovators Award to Virustatic and the Isaac Newton Institute, grant reference  10074419.

J. Houghton, L. Hope, P. Hope, I. M. Griffiths, and C. J. W. Breward contributed to the conceptualization of the problem. J.Houghton and P. Hope provided background technical expertise, while L. Hope provided legislative background. The mathematical model was formulated, solved, and interpreted by M. D. Shirley, I. M. Griffiths, and C. J. W. Breward. The write-up was completed by M. D. Shirley, I. M. Griffiths, and C. J. W. Breward.

\section{Copyright}
For the purpose of open access, the authors have applied a CC BY public copyright licence to any author-accepted manuscript (AAM) arising from this submission.

\appendix

\section{Rounding the corners of a mask}\label{sec:Rounding}

\begin{figure}[h]
    \centering
    \includegraphics[width=0.7\textwidth]{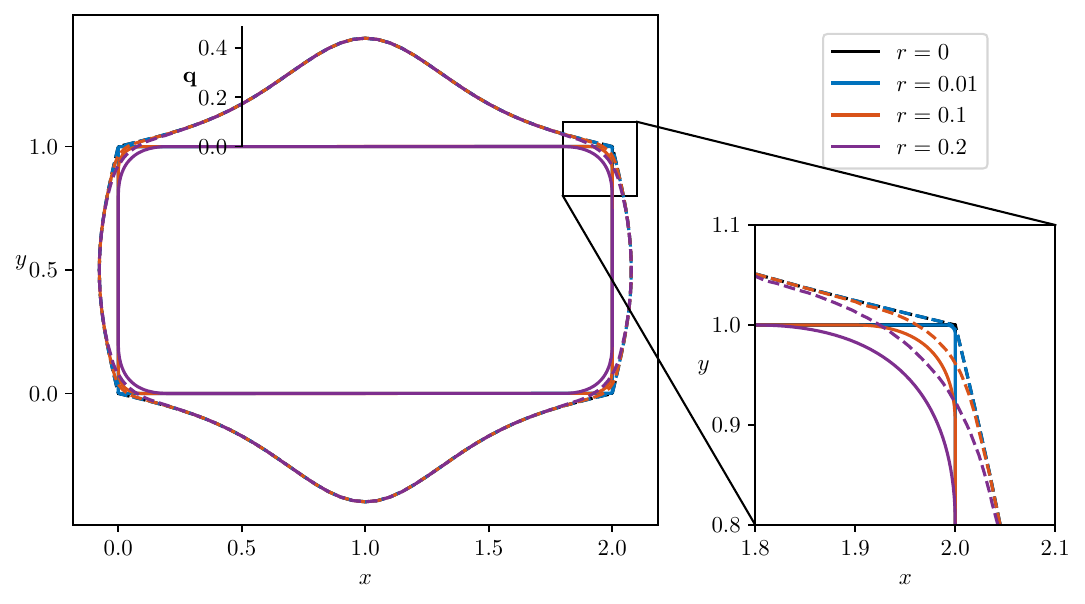}
        \caption{The effect of the sharp corners of the mask shape is investigated by plotting the mask boundary (solid lines) and the mask boundary plus the flux in the gap $\qvec$ evaluated along the boundary (dashed line) when $h\equiv1$, $k=1$, and the mouth occupies $0.852 \le x \le 1.148$, $0.475 \le y \le 0.525$.  The results with no rounding are shown in black, and results with the corner replaced by a quarter circle of radius $r$ are shown in blue, red, and purple for $r=0.01$, $r=0.1$, and $r=0.2$, respectively.  }
    \label{fig:Corners}
\end{figure}

In Fig.~\ref{fig:2D_Example_Solution_Mask}, we saw that the flux tends to zero at the corners of the mask. 
To investigate whether this effect is also present for masks with rounded corners, we solve the reduced model for $Q$, \eqref{eq:Asympt_Reduced_gov_eqn}--\eqref{eq:Asympt_Reduced_bc}, numerically using the finite element package Firedrake \cite{FiredrakeUserManual}, imposing a mask shape with rounded corners of radius $r$. 
In Fig.~\ref{fig:Corners}, we vary the amount of rounding of the mask boundary, and see that the fluxes are unaffected away from the corners, and only change appreciably in a small region close to the corners of the mask, where the flux becomes small but non-zero when the rounding is applied. This demonstrates that the flux approaching zero in the corners, while on the face of it unphysical, is only a local effect, with minimal impact on the overall behaviour of the system.

\section{Solution for a neck gaiter}\label{sec:gaiter}

We now derive the solution to \eqref{eq:Asympt_Reduced_gov_eqn} subject to the Dirichlet condition \eqref{eq:Asympt_Reduced_bc} at $y=0$ and $y=1$, and the periodic condition \eqref{eq:Asympt_Reduced_bc_periodic} at $x=0$ and $x=l$. We will assume the solution is sufficiently smooth that the periodic condition becomes
\begin{equation}
    \frac{\partial \widetilde{Q}}{\partial x} = 0 ~ \text{ at } x=0,\,x=l,
\end{equation}
where throughout we use $\widetilde{\cdot}$ notation to distinguish expressions for a neck gaiter from those for a face mask. 
Subject to these boundary conditions, the series expansion for $\widetilde{Q}$ is 
\begin{equation}
    \widetilde{Q} = \frac{1}{2}\sum_{n=1}^{\infty}\widetilde{\alpha}_{0n}\sin\left( n\pi y\right) + 
    \sum_{m=1}^{\infty}\sum_{n=1}^{\infty} \widetilde{\alpha}_{mn} \cos\left( \frac{m\pi x}{l}\right)\sin\left( n\pi y\right),\label{eq:2D_constant_series_Q_gaiter}
\end{equation}
where the coefficients $\widetilde{\alpha}_{mn}$ are given by
\begin{equation}
    \widetilde{\alpha}_{mn} =\frac{4}{l}\int_{0}^{1}\int_{0}^{l} \widetilde{Q}(x,y)\cos\left( \frac{m\pi x}{l}\right)\sin\left( n\pi y\right) \,\mathrm{d}x\mathrm{d}y= \displaystyle\frac{ k ^2 \widetilde{\beta}_{mn} }{\displaystyle \frac{m^2 \pi^2}{l^2} + n^2 \pi^2 +  k ^2 },
\end{equation}
for $m=0,1,\dotsc$ and $n=1,2,\dotsc$, where, as in the case of a face mask, $\widetilde{\beta}_{mn}$ are the coefficients of the series representation of $w_{\In}$ in this basis, {\it i.e.}
\begin{equation}
    w_{\In}(x,y) = 
    \frac{1}{2}\sum_{n=1}^{\infty} \widetilde{\beta}_{0n}\sin\left( n\pi y\right)+\sum_{m=1}^{\infty}\sum_{n=1}^{\infty} \widetilde{\beta}_{mn}\cos\left( \frac{m\pi x}{l}\right) \sin\left( n\pi y\right),
\end{equation}
and are given by
\begin{equation}
    \widetilde{\beta}_{mn} = \frac{4}{lM}\iint_{\OmegaMouth}\cos\left( \frac{m\pi x}{l}\right)\sin\left( n\pi y\right) \,\mathrm{d}x\mathrm{d}y,\label{eq:Gaiter_beta_general}
\end{equation}
for a general mouth. For a rectangular mouth with centre $\left(E_{x},E_{y}\right)$ and dimensions $2L_{x}$ and $2L_{y}$, in the $x$- and $y$-directions respectively, then
\begin{equation}
    \widetilde{\beta}_{0n}  = \frac{16}{n \pi M} \frac{L_{x}}{l}\sin \bigg(  n\pi E_{y}\bigg)\sin \bigg(  n\pi L_{y}\bigg),
\end{equation}
for $n=1,2,\dotsc$ and
\begin{equation}
    \widetilde{\beta}_{mn} = \frac{16}{mn\pi^2M}\cos\left( \frac{m\pi E_{x}}{l} \right)\sin\left( \frac{m\pi L_{x}}{l} \right)\sin \bigg(  n\pi E_{y}\bigg)\sin \bigg( n\pi L_{y}\bigg),
\end{equation}
if $m,n=1,2,\dotsc$. The depth-integrated flux of the air behind the neck gaiter is found by \eqref{eq:Asympt_sol_q}, giving
\begin{subequations}
\label{eq:2D_constant_series_q_gaiter}
\begin{align}
    \widetilde{q}_{x} &= \sum_{m=1}^{\infty} \sum_{n=1}^{\infty} \frac{m\pi}{l k ^2}\widetilde{\alpha}_{mn} \sin\left( \frac{m\pi x}{l}\right) \sin\left( n\pi y\right),\\
    \widetilde{q}_{y} &= - \frac{1}{2}\sum_{n=1}^{\infty} \frac{n\pi}{ k ^2}\widetilde{\alpha}_{0n}  \cos\left( n\pi y\right)   - \sum_{m=1}^{\infty}\sum_{n=1}^{\infty} \frac{n\pi}{ k ^2}\widetilde{\alpha}_{mn} \cos\left( \frac{m\pi x}{l}\right) \cos\left( n\pi y\right).
\end{align}
\end{subequations}
Finally, we consider the total fluxes, defined by \eqref{eq:Def_totalflux}. As we assume the solution is periodic in the $x$-direction, $\widetilde{\eta}_{\textrm{left}}=0$, $\widetilde{\eta}_{\textrm{right}}=0$. For the remaining fluxes, we find that 
\begin{equation}
    \widetilde{\eta}_{\textrm{mask}}   = \frac{1}{2}\sum_{n=1}^{\infty} \frac{l}{n\pi}\widetilde{\alpha}_{0n} \left( 1- \left(-1\right)^{n}\right),\quad ~ \widetilde{\eta}_{\textrm{bottom}} = \frac{1}{2}\sum_{n=1}^{\infty} \frac{ln\pi}{ k ^2}\widetilde{\alpha}_{0n},\quad ~
    \widetilde{\eta}_{\textrm{top}}  = -\frac{1}{2}\sum_{n=1}^{\infty} \frac{ln\pi}{ k ^2}\widetilde{\alpha}_{0n}\left(-1\right)^{n}.
\end{equation}
Note that these global quantities only depend on the zeroth mode in the $x$-direction (since all other modes integrate to zero). Using these expressions, the leakage ratio may be calculated using \eqref{eq:Def_eta}.

\section{Effect of mouth shape}\label{sec:App_shape}

In this appendix, we consider the effect of varying the shape of the mouth, $\OmegaMouth$. For simplicity, we only present results for a face mask with constant gap width, but the findings generalise to a neck gaiter, and to a face covering with variable gap width.

\begin{figure}[htbp]
    \centering
    \begin{overpic}[width=0.49\linewidth]{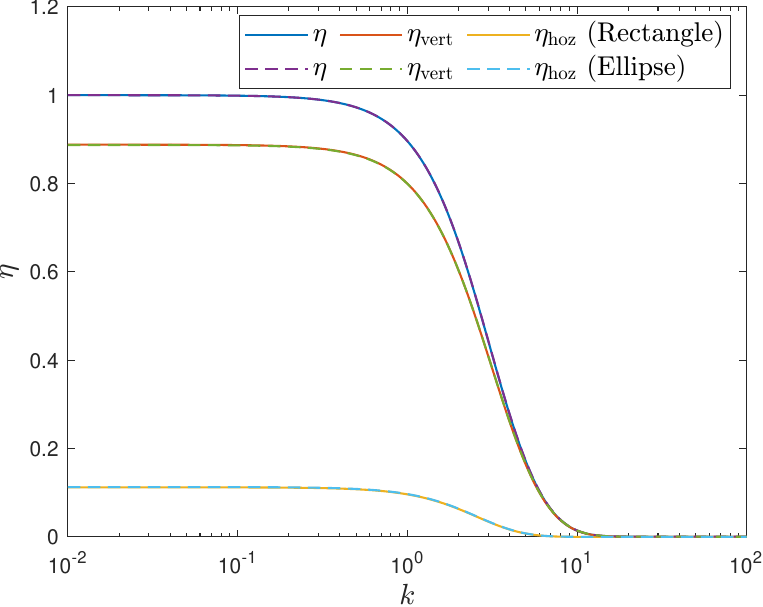} 
        \put(3,95){\textbf{(a)}}
    \end{overpic}
    \hfill
    \begin{overpic}[width=0.49\linewidth]{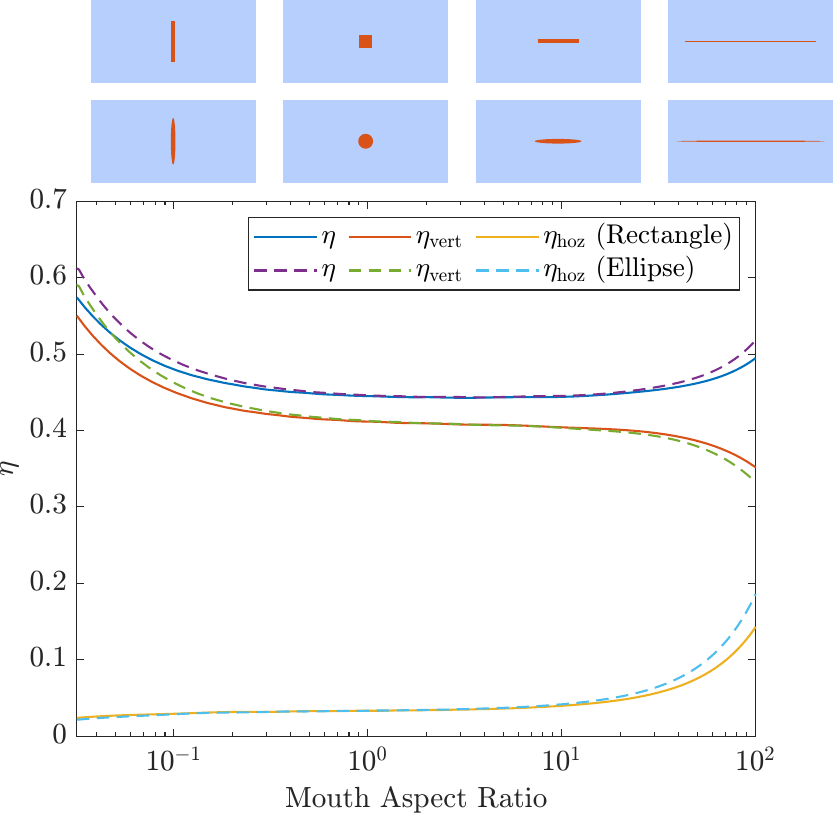}
        \put(3,95){\textbf{(b)}}
    \end{overpic}
    \caption{The total, horizontal, and vertical leakage ratios of a mask with constant gap width for a rectangular mouth (solid line) and an elliptical mouth (dashed line).  (a) The leakage ratios as functions of $ k $, keeping the aspect ratio of both mouths 2.5. (b) The leakage ratios as functions of the aspect ratio of the masks, keeping $ k =1$. The dimensionless mask width $a=2$, and the mouth is always located at the centre of the mask. }
    \label{fig:App_MouthShape}
\end{figure}

If $\OmegaMouth$ is elliptical with centre $\left(E_{x},E_{y}\right)$ and width and height $\left(2R_{x},2R_{y}\right)$, respectively,  we find that the expression for the coefficients of the series representation of $w_{\In}$, \eqref{eq:2D_constant_series_beta_general}, are 
\begin{equation}
        \beta_{mn} =  \frac{8}{l n \pi M}\sin\left( n \pi E_{y}\right)
         \int_{E_{x}-R_{x}}^{E_{x}+R_{x}}\sin\left( \frac{m\pi x}{l} \right) \sin\left( n\pi R_{y}\sqrt{1 - \left( \frac{x-E_{x}}{R_{x}} \right)^2} \right) \,\mathrm{d}x.\label{eq:2D_constant_h_beta_ellipse}
\end{equation}
The integral must be evaluated numerically, which we perform using Simpson's rule. Using this expression, we plot in Fig.~\ref{fig:App_MouthShape}a, how the total, horizontal, and vertical leakage ratios for a face mask vary with $ k $ for both a rectangular and an elliptical mouth, positioned at the centre of the mask, and with equal areas. We also keep equal the aspect ratios of the mouths, defined as their width divided by their height, i.e. $L_{x}/L_{y}$ and $R_{x}/R_{y}$ for a rectangle and an ellipse, respectively. We conclude that, at this aspect ratio, the leakage ratios are indistinguishable.

In Fig.~\ref{fig:App_MouthShape}b we again show the leakage curves for rectangular and elliptical mouths, but varying their aspect ratio and keeping $ k $ fixed. This plot shows that the leakage for the two mouth shapes are indistinguishable for moderate aspect ratios that encompass all possible realistic mouth shapes, and the leakage ratio remains approximately constant. For extremely small or large aspect ratios, the total leakage increases as the edge of the mouth becomes close to a boundary, greatly increasing the vertical or horizontal leakage, as appropriate. Since $\eta$ is increasing as the aspect ratio tends either to zero or infinity, there is also an aspect ratio that minimises the leakage. We find this to be 3.31 for a rectangular mouth and 4.07 for an elliptical mouth. Interestingly, this is slightly higher than 2, the equivalent aspect ratio of the standard face mask considered in this paper.

\bibliography{References}

\end{document}